\documentclass[showpacs,superscriptaddress,floatfix,aps,prl,notitlepage,reprint]{revtex4-2}
\usepackage{graphicx,amsfonts,amssymb,amsmath,hyperref,hypcap,enumerate}
\usepackage{braket}
\usepackage{lipsum} 
\usepackage{amsthm}
\usepackage{multirow}
\usepackage{graphicx}
\usepackage{dcolumn}   
\usepackage{bm}        
\usepackage{amssymb}   
\usepackage{amsmath}
\usepackage{braket}
\usepackage{epstopdf}
\usepackage{mathtools}

\usepackage{color}
\usepackage{subfigure}
\usepackage{diagbox}
\usepackage{longtable}
\usepackage{makecell}
\usepackage{bbold}

\usepackage{comment}

\newcommand{\ii}{\text{i}}
\newcommand{\Tr}{{\rm Tr}}
\newcommand{\mU}{\mathcal{U}}
\newcommand{\mH}{\mathcal{H}}

\newcommand{\mM}{\mathcal{M}}

\newcommand{\tf}{\tilde{f}}

\DeclareMathAlphabet{\mathitb}{OT1}{cmr}{bx}{sl}
\usepackage[utf8]{inputenc}
\usepackage{enumitem}

\usepackage{etoolbox}
\makeatletter
\patchcmd{\frontmatter@abstract@produce}
  {\vskip200\p@\@plus1fil
   \penalty-200\relax
   \vskip-200\p@\@plus-1fil}
  {}
  {}
  {}
\makeatother

\begin{document}

\title{Universal Stochastic Equations of Monitored Quantum Dynamics}

\author{Zhenyu Xiao}
\email{wjkxzy@pku.edu.cn}
\affiliation{International Center for Quantum Materials, Peking University, Beijing 100871, China}

\author{Tomi Ohtsuki}
\affiliation{Physics Division, Sophia University, Chiyoda-ku, Tokyo 102-8554, Japan}

\author{Kohei Kawabata}
\email{kawabata@issp.u-tokyo.ac.jp}
\affiliation{Institute for Solid State Physics, University of Tokyo, Kashiwa, Chiba 277-8581, Japan}

\date{\today}

\begin{abstract}
We investigate the monitored quantum dynamics of Gaussian mixed states and derive the universal Fokker-Planck equations that govern the stochastic time evolution of entire density-matrix spectra, obtaining their exact solutions.
From these equations, we reveal an even-odd effect in purification dynamics:
whereas entropy exhibits exponential decay for an even number $N$ of complex fermions, algebraic decay with divergent purification time occurs for odd $N$ as a manifestation of dynamical criticality.
Additionally, we identify the universal fluctuations of entropy in the chaotic regime,  
serving as a non-unitary counterpart of the universal conductance fluctuations in mesoscopic electronic transport phenomena.
Furthermore, we elucidate and classify the universality classes of non-unitary quantum dynamics based on fundamental symmetry.
We also validate the universality of these analytical results through extensive numerical simulations across different types of models.
\end{abstract}

  \maketitle

{\it Introduction.}---Entropy represents the uncertainty of physical systems~\cite{renyi1961}.
Information is scrambled by unitary dynamics and acquired through measurement, both of which constitute principal ingredients in quantum computation and information.
Their competition has recently been shown to induce dynamical purification phase transitions unique to open quantum systems~\cite{vasseur2019, chan2019, skinner2019, li2018a, li2019, choi2020, gullans2020d, jian2020, fisher2023a}.
Such measurement-induced phase transitions and related phenomena have been extensively studied in circuit~\cite{szyniszewski2019, bao2020, tang2020, gullans2020, zabalo2020, goto2020, lavasani2021a, sang2021, ippoliti2021a, szyniszewski2020, li2021b, fidkowski2021, nahum2021, ippoliti2021b, lunt2021, bao2021, lu2021a, doggen2022, block2022, zabalo2022, agrawal2022, yang2022, wiersema2023, tantivasadakarn2024, schomerus2022, oshima2023a, 
ravindranath2023,
liu2023, moghaddam2023a, lu2023, popperl2024a, bulchandani2024, deluca2023, gerbino2024, liu2024, mochizuki2024, noel2022, koh2023, hoke2023,
mcginley2022},
spin~\cite{lang2020, biella2021, turkeshi2021, turkeshi2022a, lavasani2023, tirrito2023, murciano2023, paviglianiti2023a, schmolke2024},
and fermionic~\cite{cao2019, nahum2020a, chen2020, alberton2021, jian2022, tang2021a, buchhold2021, muller2022, zhang2022, coppola2022a, kawabata2023a, yang2023EFT, merritt2023a, legal2023a, szyniszewski2023, jian2023, swann2023, fava2023a, loio2023, poboiko2023, poboiko2024, legal2024, yajima2024, fava2024, lunt2020, jian2021a, minato2022, altland2022, yamamoto2023, lumia2024,
ravindranath2023b}
models.

Monitored free fermions have attracted significant interest due to their rich and diverse phenomena~\cite{cao2019, nahum2020a, chen2020, alberton2021, jian2022, tang2021a, buchhold2021, muller2022, zhang2022, coppola2022a, kawabata2023a, yang2023EFT, merritt2023a, legal2023a, szyniszewski2023, jian2023, swann2023, fava2023a, loio2023, poboiko2023, poboiko2024, legal2024, yajima2024, fava2024}.
Unlike their many-body counterparts, the very existence of measurement-induced phase transitions is non-trivial.
Recently, an effective field theory akin to that for the Anderson transitions~\cite{anderson1958, abrahams1979, lee1985, beenakker1997, evers2008a} has been developed~\cite{jian2022, jian2023, fava2023a, poboiko2023}, predicting the presence (absence) of phase transitions in Majorana (complex) fermions in one spatial dimension.
However, the influence of symmetry on the distinct universality classes within monitored dynamics has remained unclear.
Furthermore, few analytical results have been obtained for microscopic models of monitored free fermions, leaving their universal characteristics still largely elusive.

In this Letter, we derive the universal Fokker-Planck equations that govern the monitored dynamics of free fermions.
We obtain their exact solutions, which describe the joint distribution of density-matrix spectra under the stochastic time evolution and encode information on all orders of R\'enyi entropy.
Building upon these equations, we uncover an even-odd effect in purification dynamics: entropy exhibits algebraic decay for an odd number $N$ of complex fermions, whereas exponential decay occurs for even $N$.
Furthermore, we identify the universal sample fluctuations of entropy, serving as a non-unitary analog of the universal conductance fluctuations in mesoscopic physics~\cite{Stone85, Altshuler85, Lee85, Imry86, Altshuler86, Lee87, Washburn86}.
We generalize these findings to enriched symmetry classes and demonstrate that the universal entropy fluctuations provide a characteristic indicator of
symmetry in the non-unitary quantum dynamics.
We validate 
these analytical results
through extensive numerical simulations across various models, confirming their universality.

{\it Monitored dynamics}.---We investigate the purification dynamics of Gaussian mixed states of $N$ complex fermions under continuous measurement.
We prepare the initial state as an un-normalized density matrix $\rho_0 = \mathbb{1}$ with maximal entropy. 
The unitary dynamics $\mU_t$ is generated by a time-dependent quadratic Hamiltonian $\mH_t$.
Meanwhile, the particle number $n_i \equiv c_i^{\dag} c_i$ ($1 \leq i \leq N$) at each site is continuously measured, corresponding to a Kraus operator~\cite{jacobs2006, wiseman2009quantum},
\begin{equation} \label{eq: mM}
 \mM_t = \exp \left\{\sum_i \left[ (n_i - \langle  n_i \rangle_t) \sqrt{\gamma} d W_t^i - 
  (n_i - \langle  n_i \rangle_t)^2 \gamma dt \right] \right\} \, ,
\end{equation}
where $\langle \cdot \rangle_t \equiv \Tr( \rho_t \cdot )/\Tr(\rho_t)$ denotes the average with the density matrix $\rho_t$ at time $t$, $\gamma$ the measurement strength, and $d W_t^i$ the standard Wiener process
satisfying $\langle dW_t^i \rangle_{E} = 0$ and $\langle d W_t^i d W_{t}^j \rangle_{E} = \delta_{ij} dt$. 
Here, $\langle \cdot \rangle_{E}$ represents the ensemble average over both Wiener process and random unitary dynamics (see below).
The un-normalized density matrix $\rho_t$ evolves by a quantum trajectory $\mM_{0:t}$,
\begin{equation} 
  \rho_t = \mM_{0:t} \mM_{0:t}^{\dag}, \quad
  \mM_{0:t} \equiv  \mM_{t} \mU_{t} \ldots \mM_{\Delta t} \mU_{\Delta t} \, .
  \label{eq:densitymatrix}
\end{equation}

The product $\mM_{0:t}$ preserves Gaussianity and is calculated by the corresponding single-particle operators~\cite{bravyi2004b}. 
We introduce a single-particle Kraus operator $M_t$
by $\mM_t c_i^{\dag} \mM_t^{-1} \equiv \sum_j c_j^{\dag} (M_t)_{ji}$, satisfying $(M_t)_{ji} = e^{\epsilon_i} \delta_{ij}$ with $\epsilon_i \equiv (2 \langle n_i\rangle_t -1 )  \gamma dt+ \sqrt{\gamma} dW_t^i$, and a single-particle unitary operator $U_t \in {\rm U}(N)$ by $\mU_t c_i^{\dag} \mU_t^{-1} \equiv \sum_j c_j^{\dag}(U_t)_{ji}$.
Owing to Gaussianity, $\rho_t$ is fully encoded in the single-particle quantum trajectory $M_{0:t} \equiv M_{t} U_{t} \ldots M_{\Delta t} U_{\Delta t}$: 
$\rho_t \propto e^{\sum_{ij} 2P_{ij} c_i^{\dag} c_j}$ with $e^{2 P} \equiv M_{0:t}M_{0:t}^{\dag}$.
The two-point correlation function is obtained as $\langle c_i^{\dag} c_j \rangle_t = (\tanh P^{\rm T} + 1)_{ij}/2$,
and the eigenvalues $2z_i$'s of $2P$, uniquely determined from the normalized density matrix $\rho_t/\Tr \rho_t$, give 
the $\alpha$-R\'enyi entropy $S_{\alpha} \equiv \left(1-\alpha \right)^{-1} 
\ln \Tr\,(\rho_t/\Tr \rho_t)^{\alpha}
= \sum_{i = 1}^N f_{s \alpha}(z_i)$ with~\cite{cheong2004}
\begin{equation} \label{eq: f_s alpha}
\begin{aligned}
f_{s \alpha}(z) \equiv
\frac{1}{1 - \alpha} \ln\left[ \frac{1}{(1 + e^{2z})^{\alpha}}+\frac{1}{(1 + e^{-2z})^{\alpha}} \right] .
\end{aligned}
\end{equation}
Specifically, $S_{2}$ 
is essentially equivalent to purity.

{\it Universal Fokker-Planck equation.}---
We model $U_t$ as a random ${\rm U}(N)$ matrix distributed uniformly in the Haar measure.
The Haar randomness enables us to capture the universal chaotic feature of the monitored dynamics,
irrespective of microscopic details.
We consider the dynamics in the infinitesimal interval $\left[ t, t+\Delta t \right]$ that renormalizes the probability distribution function $p(\{ z_n\}; t)$ of $z_n$'s.
Such an incremental change is perturbatively evaluated as~\cite{brouwer1996, supplemental}
\begin{align} \label{eq: EoM z}
  \langle \Delta z_n(t) \rangle_E
  & =  \frac{\mu_n  + \nu_n}{N + 1} \gamma \Delta t ,\\  
  \langle \Delta z_n(t) \Delta z_m(t) \rangle_E & =  \frac{1  + \delta_{mn}}{N + 1} \gamma \Delta t , \label{eq: EoM z2}
\end{align}
with
\begin{equation} \label{eq: mu nu}
\mu_n = \sum_{m \neq n} \coth(z_n-z_m) , \, \nu_n =  \sum_m  (1 + \delta_{nm}) \tanh z_m.
\end{equation}
The corresponding Fokker-Planck equation for  
$p(\{ z_n\};t)$ reads
\begin{equation} \label{eq: Fokker-Planck} 
  \frac{N + 1}{\gamma} \frac{\partial p}{\partial t} 
  = - \sum_{n=1}^N \frac{\partial  [(\mu_n + \nu_n)p] }{\partial z_n }  + 
  \frac{1}{2} 
  \sum_{m,n=1}^N \frac{\partial^2 \left[(1  + \delta_{mn}) p \right] }{\partial z_n \partial z_m} \, . 
\end{equation}
The drift terms $\mu_n$'s describe level repulsion between $z_n$'s, generally occurring in the spectra of random operators~\cite{beenakker1997}. 
In contrast, $\nu_n$'s manifest positive-feedback effect: as $z_n$'s increase, $\nu_n$'s also increase, making further increases in $z_n$'s. 
This arises from the unique nature of Born measurement.
According to Born's rule,
$\mM_t$ associated with large-$n_{\rm tot}$ measurement outcomes
is more likely to occur for larger $\langle n_{\rm tot} \rangle_t$,  
resulting in even larger $\langle n_{\rm tot} \rangle_{t + \Delta t}$. 
Meanwhile, $z_n$'s are related to the total particle number $n_{\rm tot} = \sum_{i} n_i$ [i.e., $\langle n_{\rm tot} \rangle_t = \sum_{i} (\tanh z_i(t) + 1)/2$].

Another important scenario of non-unitary dynamics 
is accompanied by postselection~\cite{jian2020, jian2023, jian2022, chen2020, tang2021a, kawabata2023a, legal2023a, swann2023}, where $\mM_t$ is applied according to prior probability instead of Born probability.
We refer to this scenario as 
forced measurement
and the one discussed earlier as 
Born measurement.
In the continuous-time description, the corresponding Kraus operator reads $(M_t)_{ij} = e^{\epsilon_i} \delta_{ij}$ with white noise $\epsilon_i \equiv \sqrt{\gamma} dW_t^i$,
and the Fokker-Planck equation
for $p_{ F}(\{ z_n\}; t)$
is
obtained similarly, 
taking the same form as Eq.~(\ref{eq: Fokker-Planck}) but with $\nu_n = 0$.

With the initial condition $\rho_0 = \mathbb{1}$, we find the exact solution $p_{F}(\{ z_n\}; t)$ to Eq.~(\ref{eq: Fokker-Planck}) as~\cite{brouwer1998, ipsen2016,  supplemental}
\begin{align} \label{eq: exact p NH}
  &p_{F}(\{ z_n\}; t) 
   = \mathcal{N}(t)  \left( \prod_{n<m} (z_n - z_m) \sinh(z_n - z_m) \right)  \nonumber \\
   &\quad \times \exp \left( -\frac{N+1}{2 \gamma t} \sum_{n,m} z_n \left( -\frac{1}{N+1} + \delta_{nm} \right) z_m \right) 
\end{align}
with a normalization constant $\mathcal{N}(t)$. 
The solution $p_{B}(\{ z_n\}; t)$ 
for the Born measurement is obtained from $p_{F}(\{ z_n\}; t)$ as
\begin{equation} \label{eq: exact p Born}
  p_{B}(\{ z_n\}; t) = e^{-\frac{N}{2} \gamma t} \left( \prod_{n} \cosh z_n \right) p_{F}(\{ z_n\}; t) \, .
\end{equation}
This connection is a manifestation of Born's rule, implying that the probability of a given quantum trajectory is proportional to $\Tr \rho_t \propto  \prod_{n} \cosh z_n$~\cite{supplemental}.
The distributions in Eqs.~(\ref{eq: exact p NH}) and (\ref{eq: exact p Born}) are invariant under $\{z_n\} \rightarrow \{-z_n\}$ owing to statistical symmetry of the dynamics under the particle-hole transformation $c_i^{\dag} \rightarrow c_i$.

Equation~(\ref{eq: Fokker-Planck}) of density-matrix spectra has an analog in quantum transport phenomena of disordered mesoscopic wires~\cite{dorokhov1982, mello1988, beenakker1997, brouwer1998, mudry1999, evers2008a}.
The Fokker-Planck equations therein universally describe the gradual changes of transmission probabilities in the spatial direction and the concomitant Anderson localization.
In contrast, Eq.~(\ref{eq: Fokker-Planck}) describes the non-unitary purification dynamics, which we elucidate in this Letter.
Significantly, Eq.~(\ref{eq: Fokker-Planck}) depends solely on the measurement strength $\gamma$, serving as a non-unitary counterpart of the one-parameter scaling~\cite{abrahams1979}.
The universality of Eq.~(\ref{eq: Fokker-Planck}) is also corroborated by the underlying ${\rm U}(R)$ non-linear sigma model (NL$\sigma$M) description~\cite{jian2022, jian2023, fava2023a, poboiko2023}.
Whereas the replica index $R \to 0$ corresponds to the forced measurement, 
as well as the Anderson localization, $R \to 1$ corresponds to the Born measurement.
Below, we clarify that this difference results in distinct purification dynamics.

{\it Purification in the long time.}---
For $\alpha > 1$ and $|z| \gg 1$, 
Eq.~(\ref{eq: f_s alpha}) reduces to $f_{s \alpha}(z) \simeq {\alpha}/{(\alpha - 1)} e^{-2 |z|}$. 
Then, the long-time decay of the entropy $S_{\alpha}$ is primarily determined by $\min |z|$, and the purification time $\tau_P$ is
\begin{equation}    \label{eq: purification time}
  \tau_P^{-1} \equiv  -\lim_{t \rightarrow \infty} \frac{\ln \langle  S_{\alpha} \rangle_E }{t} = 2 \lim_{t \rightarrow \infty} \frac{\min_n |\langle z_n \rangle_E|}{t}, 
\end{equation}
where $\lim_{t \rightarrow \infty} z_n/t \equiv \eta_n $ is a Lyapunov exponent of the quantum trajectory $M_{0:t}$.
The Lyapunov exponent is determined by analyzing $z_n$'s that maximize $p(\{z_n\},t)$ in the long-time limit $t \to \infty$, equivalent to finding mean-field solutions 
to the Fokker-Planck equations~\cite{supplemental}.

\begin{figure}[bt]
  \centering
  \includegraphics[width=1\linewidth]{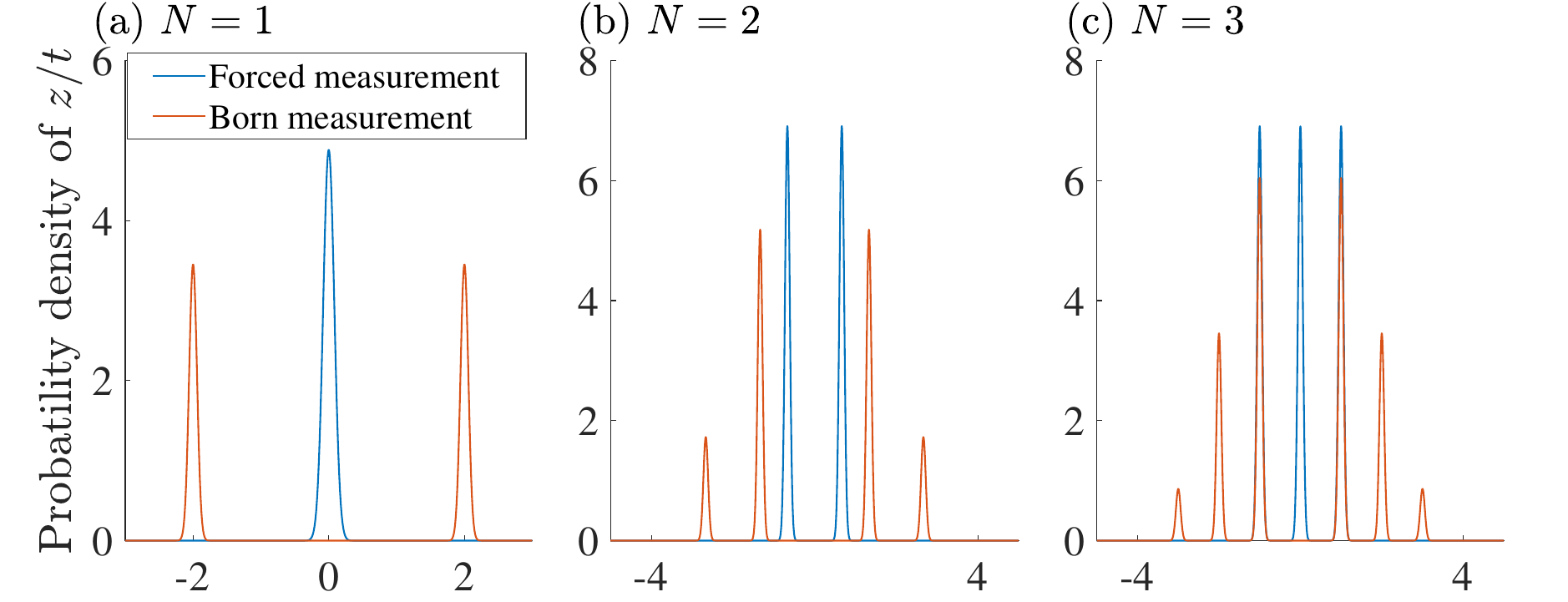}
  \caption{Distribution $\rho(z/t) = \sum_n \langle \delta(z/t - z_n/t) \rangle_E$ of single-particle density-matrix spectra for
  the exact solutions [Eqs.~(\ref{eq: exact p NH}) and (\ref{eq: exact p Born}) with $\gamma = 2$, $t = 600$, and $N = 1,2,3$] to the Fokker-Planck equation~(\ref{eq: Fokker-Planck}). 
  } 
  \label{fig: pz}
\end{figure}

\begin{figure}[tb]
  \centering
  \includegraphics[width=1\linewidth]{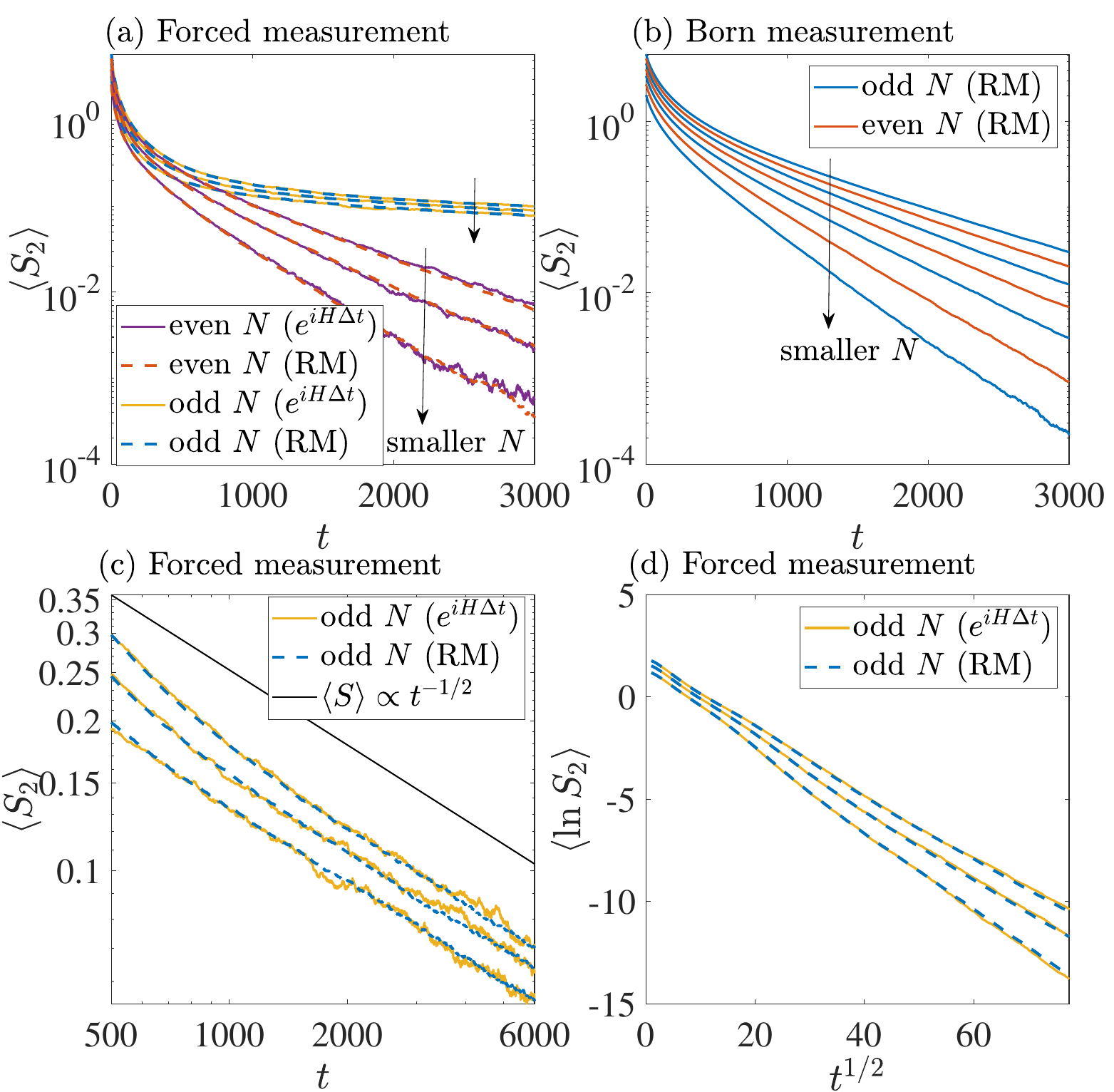}
\caption{
Numerical simulation of the long-time behavior of entropy $S_2$ for the forced and Born measurements of $N$ fermions ($\gamma = 0.16$). 
For each parity of $N$, the curves from top to down are in the descending order of $N$ ($ 9 \geq N \geq 4$), consistent with Eqs.~(\ref{eq: MF non-Hermitian}) and (\ref{eq: MF Born}).
The unitary dynamics $U_t$ is either a Haar-random ${\rm U}(N)$ matrix (RM; dashed lines) or generated by a one-dimensional local Hamiltonian ($e^{\ii H \Delta t}$; solid lines). 
See the Supplemental Material~\cite{supplemental} for details on the Hamiltonians, parameters, and algorithm.
} 
\label{fig: long_time S} 
\end{figure}

Let us order $z_n$'s by $z_1 \leq z_2 \leq \ldots \leq z_N$.
In the long-time limit, $z_n$'s should be well separated, leading to $\coth(z_n - z_m) \simeq {\rm sign}(z_n - z_m)$.
For the forced measurement,
we put this into Eqs.~(\ref{eq: EoM z}) and (\ref{eq: EoM z2}) with $\nu_n = 0$ and have
\begin{equation} \label{eq: MF non-Hermitian}
  \langle z_n  \rangle_E = \frac{2 n - N - 1}{N + 1}  \gamma t, \quad {\rm Var} (z_n) = \frac{2 \gamma t }{N+1},
\end{equation}
which are consistent with the exact solution in Eq.~(\ref{eq: exact p NH}) (see Fig.~\ref{fig: pz}).
Crucially, the purification time $\tau_P$ behaves differently depending on the parity of $N$.
We have $\min_n |\langle z_n \rangle| /t = 0$ ($\gamma/(N+1)$) and the infinite (finite) purification time $\tau_P$
for odd (even) $N$, 
further implying the algebraic (exponential) decay of entropy (Fig.~\ref{fig: long_time S}). 
For odd $N$, $z_{(N+1)/2}$ conforms to the Gaussian distribution $\varphi_G (z)$ with zero mean and variance $ 2 \gamma t/(N+1)$. 
Since $S_{\alpha}$ is mainly contributed by $z_{(N+1)/2}$, we have $\langle S_{\alpha} \rangle = \int f_{s \alpha}(z) \varphi_G(z) dz \propto t^{-1/2}$ and $\langle \ln S_{\alpha} \rangle \propto -t^{1/2}$ for $t \rightarrow \infty$.
We confirm this algebraic decay even in the non-unitary dynamics generated by a one-dimensional local Hamiltonian, showing the universality [Figs.~\ref{fig: long_time S}\,(c) and (d)].

The mean-field solutions for the Born measurement are more intricate because of non-trivial $\nu_n$.
To proceed, we assume $z_n \ll -1$ ($z_n \gg 1$) for $n \leq l$ ($n > l$) with an integer $l = 0, 1, \cdots, N$ to be determined, yielding
\begin{equation} \label{eq: MF Born}
  \langle z_n  \rangle_E = \frac{2 (n -l)- 1 + {\rm sign}(n-l-1/2) }{N + 1} \gamma t .
\end{equation} 
For any $l$, this solution satisfies the assumption 
and is self-consistent.
Thus, there exist $N+1$ distinct mean-field solutions characterized by $l = 0, 1, \cdots, N$.
Each of them represents a local maximum of $p_{B}$, and the corresponding steady state is a random pure fermionic Gaussian state with $N-l$ fermions occupied~\cite{magan2016}.
The complete distribution of $z_n$'s is their superposition, 
and the weight of the $l$th mean-field solution (i.e., probability of $z_n$'s occurring around it) is $\Tr(\mathbb{1}_{l})/\Tr(\mathbb{1}) =  C_N^l/2^N$, with the projection operator $\mathbb{1}_{l}$ to the $(N-l)$-particle subspace and the binomial coefficient $C_N^l$~\cite{supplemental}. 
This mean-field analysis is supported by the exact solution (Fig.~\ref{fig: pz}).
In contrast to the forced measurement, the Lyapunov exponent $\min_n |\eta_n| = 2\gamma/\left( N+1 \right)$ is non-vanishing for arbitrary $N$, and $\langle S_{\alpha} \rangle_E$ always decays exponentially.
This difference originates from the positive-feedback effect of $z_n$'s discussed earlier.
The purification time $\tau_P = \left( N+1 \right)/4\gamma$ linearly increases with $N$, consistent with Refs.~\cite{fidkowski2021, loio2023}.

The even-odd effect of purification is reminiscent of delocalization in coupled one-dimensional random-hopping chains~\cite{brouwer1998, mudry1999, evers2008a}. 
The localization length $\xi$ diverges only for an odd number of channels, analogous to the divergence of the purification time $\tau_P$ in the %
monitored dynamics.
While the former requires chiral symmetry, the latter does not.
The absence of the divergent purification time for the Born measurement 
should stem from the different replica index $R \to 1$ of NL$\sigma$M, which prohibits its spontaneous symmetry breaking in $0+1$ dimension.

\begin{figure}[tb]
  \centering
  \includegraphics[width=1\linewidth]{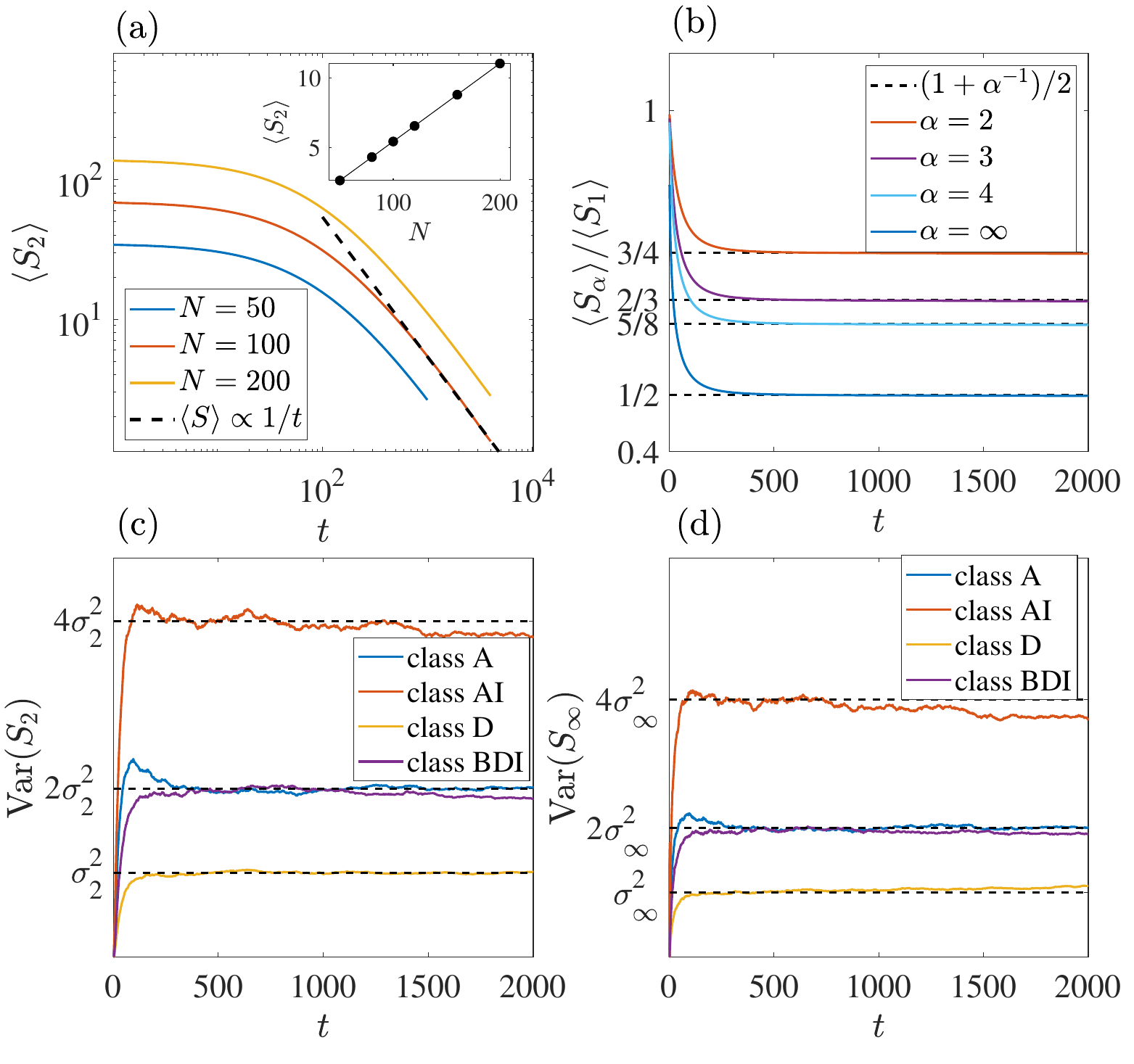}
\caption{
Numerical simulation of monitored dynamics in different symmetry classes.
(a), (b)~Entropy $\langle S_{\alpha} \rangle$ as a function of time $t$ in the dynamics of $N$
fermions [$N=200$ for (b)].
Inset of (a): $\langle S_2 \rangle$ at $t = 1000$ as a function of $N$.
(c), (d)~Variance ${\rm Var}(S_{\alpha})$ in different symmetry classes. 
The dashed lines are the analytical results ($2\sigma_2^2 =  0.06309 \ldots$ and $2\sigma_{\infty}^2 = 0.04841 \ldots$). 
The measurement strength is set to $\gamma = 0.04$ ($0.0025$) for classes A and AI (classes D and BDI);
see the Supplemental Material~\cite{supplemental} for details 
and more simulations.}  
\label{fig: mean_var_S}
\end{figure}

{\it Universal entropy fluctuations
in the short time.}---We also uncover the universal behavior in the large-$N$ and short-$t$ limit $1 \ll \gamma t \ll N$.
In both types of non-unitary dynamics, 
the spacing of two neighboring $\langle  z_n \rangle $'s is $2 \gamma t/(N+1) \ll 1$ [see Eqs.~(\ref{eq: MF non-Hermitian}) and (\ref{eq: MF Born})].
Consequently, the density $\rho(z) = \sum_n \langle \delta(z - z_n) \rangle_{E}$ is approximated as a uniform distribution: 
$\rho(z) \simeq {N}/2 \gamma t$ for $z \in [-\gamma t, \gamma t]$ and $\rho(z) \simeq 0$ otherwise~\footnote{For the Born measurement, although $\rho(z)$ is a superposition of different mean-field solutions, the weight of the $l$th solution is $C_N^l/2^N$ and hence vanishes for $N \gg 1$ and $(l-N/2)/\sqrt{N} \gg 1$, implying that the mean-field solution with $l = N/2$ suffices.}.
We then find
\begin{equation} 
    \langle S_{\alpha} \rangle_E \simeq \frac{N}{2 \gamma t} \int_{-\infty}^{ \infty} f_{s \alpha} (z)  dz = \frac{\pi^2 N}{24 \gamma t} \left(1 + \frac{1}{\alpha} \right) \, ,
\end{equation}
consistent with the numerical calculations [Figs.~\ref{fig: mean_var_S}\,(a) and (b)].
The same prefactor $1+1/\alpha$ also appears in the entanglement entropy at 
$\left( 1+1 \right)$-dimensional quantum critical points~\cite{calabrese2009, fava2023a}.

Moreover, we demonstrate the universal fluctuations of $S_{\alpha}$.
In the large-$N$ limit, we expand the variance ${\rm Var}(S_{\alpha}) \equiv \langle S_{\alpha}^2 \rangle_E - \langle  S_{\alpha}\rangle_E^2$ by $y_n \equiv z_n -  \langle z_n \rangle_{ E}$~\cite{chalker1993}.
The leading order yields ${\rm Var}(S_{\alpha})  = \sum_{i,j} f_{s \alpha}^{\prime}(\langle z_i \rangle_E)
  f_{s \alpha }^{\prime}(\langle z_j \rangle_E) \langle y_i y_j \rangle_E$. 
We also expand the distribution $p(\{z_n\};t) \equiv e^{- W(\{ z_n \};t)}$ in Eqs.~(\ref{eq: exact p NH}) and (\ref{eq: exact p Born}) around the local minimum of $W(\{ z_n \};t)$,
resulting in a Gaussian-type distribution, and subsequently evaluate $\langle y_n y_m \rangle_E$.
Performing the Fourier transformation and replacing the sum by integral in the expansion of ${\rm Var}(S_{\alpha})$, we obtain
\begin{equation} 
  {\rm Var}(S_{\alpha})  =   \int_{-\infty}^{\infty} dq  \frac{|q| (1 - e^{- \pi |q| })}{4 \pi^2} \tf_{s \alpha}(q)^2, 
\end{equation}
with $\tf_{s \alpha} (k) \equiv \int_{-\infty}^{\infty} f_{s \alpha}(z) e^{-\ii k z} dz$.
Thus, ${\rm Var}(S_{\alpha})$ yields a remarkably universal constant for both types of non-unitary dynamics, similar to the universal conductance fluctuations in mesoscopic physics~\cite{Stone85, Altshuler85, Lee85, Imry86, Altshuler86, Lee87, Washburn86}.
Specifically, for  
$S_2$, we have $\tf_{s2} (k) = \pi  \tanh \left({\pi  k}/{8}\right)/\left( k \cosh \left({\pi  k}/{4}\right) \right)$ and hence ${\rm Var}(S_2) = 2 \sigma_2^2 \equiv 10 \ln 2 -6 \ln \pi = 0.06309 \ldots$. 
We confirm the universality by simulating the unitary dynamics $U_t$ by a local Hamiltonian instead of the Haar-random matrix [Figs.~\ref{fig: mean_var_S}\,(c) and (d)]~\cite{supplemental}.
The universal entropy fluctuations arise even for projective measurement, which cannot be directly described by our Fokker-Planck equations.

{\it Symmetry classification}.---Symmetric space of the quantum trajectory $M_{0:t}$
greatly influences the time evolution of its singular-value spectrum, as also noticed in the study of quantum transport~\cite{beenakker1997, evers2008a}.
In the dynamics studied above, the unitary part $U_t \in {\rm U}(N)$ imposes no symmetry constraint on $M_{0:t}$. 
Therefore, the dynamical generator $L_{\rm eff}$ defined by $M_{0:t} \equiv e^{L_{\rm eff} t}$ is a generic non-Hermitian matrix without any symmetry and hence belongs to class A (see Table~\ref{tab: sym class})~\cite{altland1997a, Bernard-LeClair-02, kawabata2019}.

As an exemplary symmetry class different from class A, we study
the monitored dynamics of $2N$ Majorana fermions
~\cite{fava2023a}. 
A generic Majorana quadratic Hamiltonian is $\mH = \sum_{i,j} H_{ij} \psi_i \psi_j$ ($H^{\dag} = H$, $H^{\rm T} = -H$) with Majorana fermions $\psi_i$'s ($\psi_i = \psi_i^{\dag}$, $\{ \psi_i, \psi_j \} = 2 \delta_{i,j}$).
Gaussian Majorana unitary operators satisfy
$e^{-\ii \mH \Delta t} \psi_i e^{\ii \mH \Delta t} = \sum \psi_j (U_t)_{ji}$ and $U_t = e^{-4 \ii H \Delta t} \in {\rm SO}(2N)$, 
and Gaussian measurements satisfy $M_t^{\rm T} = M_t^{-1}$.
Consequently, the quantum trajectory $M_{0:t}$, comprised of the product of $U_t$'s and $M_t$'s, satisfies symmetry $M_{0:t}^{\rm T} = M_{0:t}^{-1}$, and hence the non-Hermitian dynamical generator respects $L_{\rm eff}^{\rm T} = -L_{\rm eff}$ and belongs to class D~\cite{altland1997a, Bernard-LeClair-02, kawabata2019}. 
Due to this symmetry, the singular values of $M_{0:t}$ come in $(e^{-z_n}, e^{z_n})$ pairs ($z_n \geq 0$), leading to 
\begin{align} \label{eq: EoM z D}
  \langle \Delta z_n(t) \rangle_E 
   &=   \frac{4 \left(  \mu_n  + \nu_n \right)}{2N - 1} \gamma \Delta t  ,\\  
  \langle \Delta z_n(t) \Delta z_m(t) \rangle_E  &=  \frac{4 \delta_{mn}}{2 N-1} \gamma \Delta t  , \label{eq: EoM z2 D}
\end{align}
with
\begin{equation} 
  \mu_n = \sum_{m \neq n} \left( \coth(z_n - z_m) + \coth(z_n + z_m) \right) , \, \nu_n = \tanh z_n \,.
\end{equation}
We obtain the exact solutions to the corresponding Fokker-Planck equations and find the universal entropy fluctuations in the short time, ${\rm Var}(S_{\alpha}) = \sigma_{\alpha}^2$, 
half of those in class A [Figs.~\ref{fig: mean_var_S}\,(c) and (d)]~\cite{supplemental, vidal2003}.
Similar to disordered spinful superconductors~\cite{brouwer2000}, 
the algebraic purification under forced measurement arises for arbitrary $N$, also implying a distinct universality class from class A.

\begin{table}[tb]
  \centering
  \caption{Symmetry classification of non-unitary quantum dynamics. 
  The column ``U(1)'' specifies whether quadratic Hamiltonians for unitary dynamics respect U(1) symmetry. 
  The columns ``$H_t$'', ``$M_{0:t}$'', and ``$L_{\rm eff}$'' specify the symmetry class of $H_t$, symmetric space of quantum trajectories
  $M_{0:t}$, and symmetry class of non-Hermitian dynamical generators $L_{\rm eff}$, respectively. 
  If static disordered Hamiltonians belong to the class in the column ``$H_{\rm dis}$'', their transfer matrices belong to the same symmetric space as that of $M_{0:t}$.}
  \begin{tabular}{ccc ccc}
  \hline \hline
  U(1) & ~$H_t$~& $M_{0:t}$ & ~$L_{\rm eff}$~ & ~$H_{\rm dis}$~& ~${\rm Var}(S_{\alpha})$~ \\ \hline
  $\surd$ & A   & ${\rm GL}(N,\mathbb{C})/{{\rm U}(N)}$ & A & AIII & $2 \sigma_{\alpha}^2$ \\
  $\surd$  & D   &  ${\rm GL}(N,\mathbb{R})/{\rm O}(N)$ & AI & BDI & $4 \sigma_{\alpha}^2$ \\
  $\times$ & D  &  ${\rm SO}(2N,\mathbb{C})/{\rm O}(2N)$  & D & DIII &$ \sigma_{\alpha}^2$ \\
  $\times$ & D$\oplus$D  &  ${\rm O}(N,N)/{\rm O}(N) \times {\rm O}(N)$ & BDI & D & $2 \sigma_{\alpha}^2$ \\ \hline
  \end{tabular}
  \label{tab: sym class}
\end{table}

Symmetry of non-unitary dynamics is further enriched if the Hamiltonian $H_t$ respects additional symmetry.
For monitored complex fermions, particle-hole symmetry of $H_t$ (i.e., $H_t^{\rm T} = -H_t$) leads to $U_t = e^{- \ii H_t \Delta t} \in {\rm SO}(N)$
and the reality constraints $M_{0:t}^{*} = M_{0:t}$, $L_{\rm eff}^* = L_{\rm eff}$, further resulting in class AI.
For monitored Majorana fermions, the block-diagonalized structure of $H_t$ (i.e., $\sigma_z H_t \sigma_z = H_t$ and $H_t^{\rm T} = - H_t$) leads to $\sigma_z U_t \sigma_z = U_t$ besides $U_t \in {\rm SO}(2N)$.
Consequently, we have $\sigma_z M_{0:t}^* \sigma_z = M_{0:t}$, $M_{0:t}^{\rm T} = M_{0:t}^{-1}$,
as well as $L_{\rm eff}^{\rm T} = - L_{\rm eff}$, $ \sigma_z L_{\rm eff}^* \sigma_z = L_{\rm eff}$, resulting in class BDI. 
In Table~\ref{tab: sym class}, we summarize the symmetry classification, determining the Fokker-Planck equations and concomitant purification dynamics (see also Fig.~\ref{fig: mean_var_S}).

{\it 
Discussion}.---We establish the Fokker-Planck equations that universally govern the monitored dynamics of free fermions.
Our formula~(\ref{eq: purification time}) relating the purification time $\tau_P$ to the Lyapunov exponents $\eta_n$'s facilitates efficient numerical analysis of measurement-induced phase transitions~\cite{unpublished}.
Due to the single-particle nature, it differs from those in Refs.~\cite{gullans2020d, zabalo2022, mochizuki2024}, and the resulting Fokker-Planck equation~(\ref{eq: Fokker-Planck}) also contrasts with many-body formulations~\cite{schomerus2022, bulchandani2024, deluca2023, gerbino2024}.
We elucidate that this distinction enriches the monitored quantum dynamics, leading to the even-odd effect of the purification dynamics.
Moreover, we uncover the universal entropy fluctuations within the chaotic regime, serving as a non-unitary counterpart of the universal conductance fluctuations~\cite{Stone85, Altshuler85, Lee85, Imry86, Altshuler86, Lee87, Washburn86}.
It warrants further investigation to 
incorporate the full-counting statistics of other observables.
In this respect, the charge fluctuations should also exhibit similar behavior since they are of the same order as R\'enyi entropy for fermionic Gaussian states~\cite{potter2014, burmistrov2017}.
Application of our approach to quantum control and engineering~\cite{wiseman2009quantum} also deserves further research.

{\it Acknowledgment.}---We thank Ryuichi Shindou for helpful discussion.
Z.X thanks Zongping Gong for comments on this work. 
Z.X. is supported by the National Basic Research Programs of China (No.~2019YFA0308401) and by the National Natural Science Foundation of China (No.~11674011 and No.~12074008).
T.O. is supported by MEXT KAKENHI Grant-in-Aid for Transformative
Research Areas A ``Machine learning physics" No.~22H05114.
K.K. is supported by MEXT KAKENHI Grant-in-Aid for Transformative
Research Areas A ``Extreme Universe" No.~24H00945.
\bibliography{ref}

\end{document}


\setcounter{equation}{0}
    \setcounter{section}{0}
    \setcounter{figure}{0}
    \setcounter{table}{0}
    \setcounter{page}{1}
    \renewcommand{\theequation}{S\arabic{equation}}
    \renewcommand{\thesection}{\Roman{section}}

    \renewcommand{\thefigure}{S\arabic{figure}}
    \renewcommand{\thetable}{\arabic{table}}
    \renewcommand{\tablename}{Supplementary Table}
    
    \renewcommand{\bibnumfmt}[1]{[S#1]}
    \renewcommand{\citenumfont}[1]{#1}

\title{Supplemental Material for \\ ``Universal Stochastic Equations of Monitored Quantum Dynamics"}
    
\author{Zhenyu Xiao}
\email{wjkxzy@pku.edu.cn}
\affiliation{International Center for Quantum Materials, Peking University, Beijing 100871, China}

\author{Tomi Ohtsuki}
\affiliation{Physics Division, Sophia University, Chiyoda-ku, Tokyo 102-8554, Japan}

\author{Kohei Kawabata}
\email{kawabata@issp.u-tokyo.ac.jp}
\affiliation{Institute for Solid State Physics, University of Tokyo, Kashiwa, Chiba 277-8581, Japan}
\date{\today}
	
\maketitle

This Supplemental Material is organized as follows.
In Sec.~\ref{sec: FP}, we derive the Fokker-Planck equations governing the stochastic time evolution of the density-matrix spectra in non-unitary dynamics across different symmetry classes. 
We derive the Fokker-Planck equation for the Born measurement of complex fermions in detail and provide its exact solution. 
In Sec.~\ref{sec: UEF}, we demonstrate that mean-field solutions to the Fokker-Planck equations correspond to the local maximum of the distribution $p(\{ z_n\};t)$ for $t \gg 1$. 
We calculate the probability weights of different mean-field solutions and evaluate ${\rm Var}(S_{\alpha})$ using two independent methods.  
In Sec.~\ref{sec: numeric}, we discuss the numerical algorithm for simulating weak and projective measurements and provide additional numerical simulations with detailed descriptions.

\tableofcontents

\section{Fokker-Planck equations in different symmetry classes}
\label{sec: FP}

\subsection{Monitored dynamics of complex fermions}
\label{subsec: MD A}

We derive the Fokker-Planck equation describing the stochastic time evolution of the singular values $e^{z_n(t)}$'s of the single-particle quantum trajectory $M_{0:t} \equiv M_{t} U_{t} \cdots M_{\Delta t} U_{\Delta t}$ for complex fermions under Born measurement.
The dynamics is discretized in the following manner:
the unitary dynamics $U_{n \Delta t} $ is applied in the interval $((n-1) \Delta t, n \Delta t )$ with $n \in \mathbb{Z}$, and the measurement $M_{n \Delta t}$ is imposed at time $n \Delta t$. 
We use $\rho_{t - 0}$ ($t \equiv n \Delta t$) to refer to the density matrix at time $t$ but before the measurement $M_{t}$, and $\rho_{t}$ to refer to that after $M_{t}$.
Here, $U_t$ is modeled as a random ${\rm U}(N)$ matrix uniformly distributed according to the Haar measure.
Additionally, $U_t$'s at different $t$ are independent and do not depend on the history of measurement results or the density matrix $\rho_{t-\Delta t}$ before the operation.
The Born measurement corresponds to
\begin{equation}  \label{eq: app M_t}
(M_t)_{ji} = e^{\epsilon_i} \delta_{ij}, \quad 
\epsilon_i 
\equiv 
\left(2 \langle n_j\rangle_{t - 0} - 1\right) \gamma \Delta t + \sqrt{\gamma} \Delta W_t^i.
\end{equation}
Notably, $M_{t}$ does not depend on the density matrix $\rho_{t}$ but on $\rho_{t-0}$ before the measurement, required by the causality.
As discussed in the main text, $2P$ defined by $M_{0:t}M_{0:t}^{\dag} \equiv e^{2 P}$ can be considered as the parent Hamiltonian of the density matrix $\rho_t$ at time $t$, such that $\rho_t \propto e^{\sum_{ij} 2P_{ij} c_i^{\dag} c_j}$. 
The two-point correlation function is given by~\cite{cheong2004} 
\begin{equation} \label{eq: app corr}
    \langle c_i^{\dag} c_j \rangle_t = \frac{1}{2} \left(\tanh P^{\rm T} + 1\right)_{ij}.
\end{equation}

The singular-value decomposition yields $M_{0:t} M_{0:t}^{\dag} = V_t \Lambda_t^2 V_t^{\dag} $ with a diagonal matrix $(\Lambda_t)_{ii} = e^{2 z_i(t)}$ and a unitary matrix $V_t$.
At $t + \Delta t$, we have $M_{0:t + \Delta t} M_{0:t + \Delta t}^{\dag}= M_{t + \Delta t} U_{t + \Delta t}   V_t \Lambda_t^2 V_t^{\dag}  U_{t + \Delta t}^{\dag}M_{0:t + \Delta t}^{\dag} $.
Since $U_{t + \Delta t}$ is independent of $M_t$ and $V_t$, and the Haar measure is invariant under multiplication, $U \equiv U_{t + \Delta t} V_t$ is also distributed uniformly in the Haar measure and independent of $V_t$.
Notably, $U$ diagonalizes $M_{0:t + \Delta t-0}$ and depends on time $t$, although we do not explicitly put a subscript to emphasize its time dependence for simplicity of notation.
$M_{0:t + \Delta t} M_{0:t + \Delta t}^{\dag}$ shares the same spectrum as $(U^{\dag} M_{t + \Delta t} U )\Lambda_t^2  (U^{\dag} M_{t + \Delta t} U )^{\dagger}$.
Let us define $w \equiv U^{\dag} M_{t + \Delta t} U - 1$, satisfying $w_{mn} = \sum_i U_{im}^* U_{in} \xi_i $ with $\xi_i \equiv (M_{t + \Delta t})_{ii} - 1$.
Replacing $t$ in Eq.~(\ref{eq: app M_t}) by $t + \Delta t$ and putting it into $\xi_i$, we have
\begin{align} 
     \xi_i &\equiv (M_{t + \Delta t})_{ii} - 1 \nonumber \\
     &= \epsilon_i + \frac{1}{2}\epsilon_i^2 + \mathcal{O}((\Delta t)^{3/2}) \nonumber \\
    &= (2 \langle  n_i \rangle_{t+\Delta t -0} -1 )\gamma \Delta t  + \sqrt{\gamma} \Delta W_{t+\Delta t}^i + \frac{1}{2} \gamma \Delta t  + \mathcal{O}((\Delta t)^{3/2}).  
\end{align}
Here, $\langle  n_i \rangle_{t+\Delta t -0}$ is determined by $M_{0:t+\Delta t-0}M_{0:t+\Delta t-0}^{\dag} = U \Lambda_t U^{\dag}$ through Eq.~(\ref{eq: app corr}) with $P = U \ln (\Lambda_t) U^{\dag}$.
Thus, we have
\begin{align} \label{eq: xi_i}
      \xi_i = \gamma \Delta t \sum_j |U_{ij}|^2 \tanh(z_j(t))  + \sqrt{\gamma}  \Delta W_{t+\Delta t}^i + \frac{1}{2} \gamma \Delta t + \mathcal{O}(\Delta t^{3/2}).
\end{align}
Using the second-order perturbation theory, the eigenvalues $e^{2 z_n(t + \Delta t)}$ of $(1 + w) \Lambda_t^2 (1 + w^{\dag})$ are given by
\begin{equation} 
  e^{2 z_n(t + \Delta t)} = e^{2 z_n(t)} + 
  2 w_{nn} e^{2 z_n(t)} + \sum_m |w_{nm}|^2 e^{2 z_m(t)} + \sum_{m \neq n} \frac{|w_{nm}|^2 ( e^{2 z_n(t)} + e^{2 z_m(t)})^2}{e^{2 z_n(t)} - e^{2 z_m(t)}} + \mathcal{O}(\Delta t^{3/2}) \, ,
\end{equation}
which leads to
\begin{equation} \label{eq: app delta z_n}
  2 z_n(t + \Delta t) - 2z_n(t) =  2 w_{nn} + \sum_m |w_{nm}|^2 +   4 \sum_{m \neq n} \frac{|w_{nm}|^2  e^{2 z_m(t)}}{e^{2 z_n(t)} - e^{2 z_m(t)}} - \frac{1}{2} (2 w_{nn})^2 + \mathcal{O}(\Delta t^{3/2}) .
\end{equation}

For each term in Eq.~(\ref{eq: app delta z_n}), we perform the ensemble average $\langle \cdot \rangle_E$ over both the Haar measure and the Wiener process.
Putting Eq.~(\ref{eq: xi_i}) in $w_{mn} = \sum_i U_{im}^* U_{in} \xi_i$, we have
\begin{equation}
    w_{nn} =  \gamma \Delta t \sum_{i,j} |U_{in}|^2 |U_{ij}|^2 \tanh(z_j(t))  + \sqrt{\gamma}  \sum_{i} |U_{in}|^2 \Delta W_{t+\Delta t}^i + \frac{1}{2} \sum_{i} |U_{in}|^2 \gamma \Delta t + \mathcal{O}(\Delta t^{3/2}).
\end{equation}
While $\Delta W_{t + \Delta t}^i$ appears in the measurement $M_{t + \Delta t}$ on $\rho_{t + \Delta t - 0}$, the influence of $\rho_{t + \Delta t - 0}$ on $M_{t + \Delta t}$ is only manifested in $\langle n_j\rangle_{t + \Delta t -0}$. 
Thus, $\Delta W_{t + \Delta t}$ is independent of $\rho_{t + \Delta t -0}$ and $U$, implying $\langle |U_{in}|^2 \Delta W_{t+\Delta t}^i \rangle_E = \langle |U_{in}|^2 \rangle_E \langle \Delta W_{t+\Delta t}^i \rangle_E = 0$.
Furthermore, with the help of the identity for ${\rm U}(N)$ random matrices~\cite{brouwer1996}, 
\begin{equation}
\langle |U_{ij}|^2 \rangle_E  = \frac{1}{N}, \quad \langle |U_{in}|^2 |U_{ij}|^2  \rangle_E = \frac{1 + \delta_{nj}}{N(N+1)},
\end{equation}
we have 
\begin{align} 
     \langle w_{nn} \rangle_E 
     &= \gamma \Delta t\sum_{i,j} \langle |U_{in}|^2 |U_{ij}|^2 \rangle_E \tanh(z_j(t)) +  \frac{1}{2} \gamma \Delta t \sum_i\langle |U_{in}|^2 \rangle_E \nonumber \\
     &= \gamma \Delta t \sum_{i,j}  \frac{(1 + \delta_{nj}) \tanh(z_j(t))}{N(N+1)}  + \frac{1}{2} \gamma \Delta t \nonumber \\
     &= \left( \frac{1}{N + 1} \tanh(z_n(t)) +  \frac{1}{N + 1} \sum_j \tanh(z_j(t)) + \frac{1}{2} \right) \gamma \Delta t.
\end{align}
Similarly, owing to the independence between $U$ and $\Delta W_t^i$, we have
\begin{align}
      \langle  w_{nn} w_{mm}\rangle_E 
      &= \sum_{i,j}  \langle U_{in}^* U_{in}  U_{jm}^* U_{jm}  \xi_i \xi_j \rangle_E \nonumber \\
      &= \gamma \Delta t  \sum_{i,j}\langle U_{in}^* U_{in}  U_{jm} U_{jm}^*  \rangle_E \delta_{ij} \nonumber \\
      &= \frac{1}{N + 1}(1  + \delta_{mn}) \gamma \Delta t, \\
     \langle  |w_{mn}|^2 \rangle_E 
     &= \sum_{i,j} \langle U_{im}^* U_{in}  U_{jm} U_{jn}^*  \xi_i \xi_j \rangle_E \nonumber \\
     &= \sum_{i,j} \langle U_{im}^* U_{in}  U_{jm} U_{jn}^*   \rangle_E \gamma \Delta t \delta_{ij} \nonumber \\  
     &= \frac{1}{N + 1}(1  + \delta_{mn}) \gamma \Delta t \, .
\end{align}
From these results, we find that the changes of $z_n$'s satisfy [\SDE]
\begin{align} \label{eq: app EoM z}
  \langle \Delta z_n(t) \rangle_E & =  \left[   \sum_j  \frac{1 + \delta_{nj}}{N+1} \tanh(z_j(t)) + 1 + \sum_{m \neq n} \frac{\coth(z_n-z_m) - 1}{N + 1} - \frac{2}{N+1}  \right] \gamma \Delta t
  =  \frac{\mu_n  + \nu_n}{N + 1} \gamma \Delta t ,\\  
  \langle \Delta z_n(t) \Delta z_m(t) \rangle_E & =  \frac{1  + \delta_{mn}}{N + 1}\gamma \Delta t \, ,  \label{eq: app EoM z2}
\end{align}
with
\begin{equation} \label{eq: app mu nu}
\mu_n \equiv \sum_{m \neq n} \coth(z_n-z_m) , \quad \nu_n \equiv  \sum_m  (1 + \delta_{nm}) \tanh(z_m) .
\end{equation}
These results lead to the following Fokker-Planck equation:
\begin{equation} \label{eq: app FP A0}
  \frac{N + 1}{\gamma} \frac{\partial p}{\partial t} 
  = - \sum_{n=1}^N \frac{\partial  \left[ (\mu_n + \nu_n )p \right] }{\partial z_n }  + \frac{1}{2} \sum_{m,n=1}^N \frac{\partial^2 \left[(1  + \delta_{mn}) p \right] }{\partial z_n \partial z_m} \, .
\end{equation}

\subsection{Forced measurement}

In the formalism of continuous measurement, the Kraus operator $\mM_t$ for the measurement on $n_i$ is a function of continuous variables $\beta_i$ (see also Sec.~\ref{subsec: exact solution}): 
\begin{equation}
    \mM_t(\beta_i)  = \left( \frac{2 \gamma \Delta t}{\pi} \right)^{1/4} e^{- \gamma \Delta t (n_i - \beta_t^i )^2 } \propto 
     e^{ \gamma \Delta t    (2 \beta_t^i - 1) n_i   } .
\end{equation}
Here, $\beta_i$ characterizes the measurement outcomes on $n_i$: when $\beta_i$ is larger, $\mM_t(\beta_i)$ has a larger component in the $n_i=1$ subspace; 
for $\beta_i = 1/2$, $\mM_t(\beta_i)$ has the equal components in the $n_i= 0$ and $n_i = 1$ subspaces. 
For Born measurement, we have $\beta_i \sim N[\langle n_i \rangle, 1/4 \gamma \Delta t]$, where $N[\mu ,\sigma^2]$ denotes the Gaussian distribution with a mean value $\mu$ and variance $\sigma^2$~\cite{jacobs2006}.
The variable $\epsilon_i$ in $M_t$ [Eq.~(\ref{eq: app M_t})] is determined as $ \epsilon_i = \gamma \Delta (2 \beta_t^i - 1)$ and hence satisfies $ \epsilon_i \sim N [ 2 \langle n_i \rangle -1, \gamma \Delta t ]$.

For forced measurement, we post-select measurement outcomes $\beta_i$, i.e., discard some measurement results. 
We require that after the post-selection, the distribution of $\beta_i$ does not depend on $ \langle n_i \rangle$; specifically, it satisfies $N[1/2, 1/4 \gamma \Delta t ]$.
Notably, it is always possible to discard random variables in the distribution $N[\langle n_i \rangle, 1/(4 \gamma \Delta t)]$ such that the remaining variables satisfy $N[1/2, 1/(4 \gamma \Delta t) ]$.
Correspondingly, we have $\epsilon_i \sim N[0 , \gamma \Delta t]$ and hence 
\begin{equation} \label{eq: app forced meas}
    (M_t)_{ji} = e^{\epsilon_i} \delta_{ij}, \quad \epsilon_i \equiv \sqrt{\gamma} \Delta W_t^i.
\end{equation}

We can also consider the case where the measurement is continuous, but the Kraus operators $\mM_t$ only depend on the two-valued outcome $s_i = \pm 1$ (see also Sec.~\ref{subsec: weak m}): 
\begin{equation} 
 \mM_{t}(s_i) = \frac{e^{ s_i \sqrt{\gamma \Delta t}  (n_i - \frac12)}}{\sqrt{2 \cosh(\sqrt{\gamma \Delta t})}} .
\end{equation}
For Born measurement, the probability $p_{\pm}$ of the measurement outcome $s_i = \pm$ depends on $\langle n_i \rangle$ [Eq.~(\ref{eq: app p cosh})]. 
For forced measurement, we discard some measurement outcomes such that the remaining ones satisfy $p_+ = p_- = 1/2$.
In such a scheme, in $M_t$ [Eq.~(\ref{eq: app M_t})], $\epsilon_i = s_i \sqrt{\gamma \Delta t}$ satisfies the binomial distribution with the mean value $0$ and variance $ \gamma \Delta t$, similar to the case of the continuous Kraus operator in Eq.~(\ref{eq: app forced meas}).

Following the same procedure as before, we find that the resulting Fokker-Planck equation with $M_t$ in Eq.~(\ref{eq: app forced meas}) is given by Eq.~(\ref{eq: app FP A0}) with $\nu_n = 0$.

\subsection{Exact solution to the Fokker-Planck equation}
\label{subsec: exact solution}
We investigate the solution $p_F(\{ z_n\};t)$ (i.e., probability distribution function of $z_n$'s) to the Fokker-Planck equation~(\ref{eq: app FP A0}) for the forced measurement. 
We change the variables in Eq.~(\ref{eq: app FP A0}) with $\nu_n = 0$ as follows:
\begin{equation}
 y_n \equiv \sum_{m} A_{nm} z_m \quad \text{with} \quad A_{nm} \equiv \frac{1}{\sqrt{N+1}} \left(\frac{1}{N} - \frac{\sqrt{N+1}}{N}\right) + \delta_{nm}, \quad s \equiv \frac{\gamma t}{N+1},
\end{equation}
satisfying $(A^{-2})_{mn} = 1 + \delta_{mn}$. 
After this transformation, Eq.~(\ref{eq: app FP A0}) with $\nu_n = 0$ reduces to
\begin{equation} \label{eq: app DMPK AIII}
  \frac{\partial p}{\partial s} = -\sum_{n=1}^N \frac{\partial}{\partial y_n} \left[ \sum_{m \neq n} \coth(y_n - y_m) \, p \right] + \frac{1}{2} \sum_{n=1}^N \frac{\partial^2 p}{\partial y_n^2},
\end{equation}
which is identical to the Fokker-Planck equation describing the gradual changes in transmission probabilities along the spatial direction of disordered mesoscopic wires~\cite{brouwer1998}.
The exact solution to Eq.~(\ref{eq: app DMPK AIII}) with the initial condition $p(\{y_i\}; s = 0) = \delta(y_1) \delta(y_2) \ldots \delta(y_N)$ is 
\begin{equation}
  p(\{y_i\}; s) = \frac{1}{(2 \pi)^{N/2} s^{N^2/2} \prod_{n=1}^{N-1} n!} e^{-\frac{N(N^2-1)}{6}s} \left( \prod_{j<k} (y_j - y_k) \sinh(y_j - y_k) \right) e^{-\frac{1}{2s} \sum_{j=1}^N y_j^2} \, .
\end{equation}
By reverting the variables, the solution to Eq.~(\ref{eq: app FP A0}) with $\nu_n = 0$ and the initial condition $\rho_0 = \mathbb{1}$ is
\begin{equation}  \label{eq: app p_RN}
\begin{aligned}
  p_F(\{z_i\}; t) = \frac{(N+1)^{N^2/2-1/2}}{(2 \pi)^{N/2} (\gamma t)^{N^2/2} \prod_{n=1}^{N-1} n!} e^{-\frac{N(N-1) \gamma}{6} t} \left( \prod_{j<k} (z_j - z_k) \sinh(z_j - z_k) \right) e^{-\frac{N+1}{2 \gamma t} \sum_{i,j} z_i \left(-\frac{1}{N+1} + \delta_{ij}\right) z_j} \, .
\end{aligned}
\end{equation}

Next, we demonstrate that the solution $p_{B} (\{ z_n\}; t)$ to the Fokker-Planck equation for the Born measurement [i.e., Eq.~(\ref{eq: app FP A0}) with $\nu_n \neq 0 $] under the initial condition $\rho_0 = \mathbb{1}$ satisfies
\begin{equation} \label{eq: app p_MD p_RN}
  p_{B}(\{ z_n\}; t)  \propto \left( \prod_{n} \cosh(z_n) \right) p_{F}(\{ z_n\}; t) \, , 
\end{equation}
which is established through (i)~an argument based on the underlying physical models and (ii)~straightforward calculations.

In the formalism of continuous measurement
(see, for example, Ref.~\cite{jacobs2006}), the Kraus operator is a function of continuous real variables ${\bm \beta_t} = (\beta_1, \beta_2,\ldots, \beta_N)$,
\begin{equation} \label{eq: app mM}
  \mM( {\bm \beta_t} ) = \left( \frac{2 \gamma \Delta t}{\pi} \right)^{N/4} \exp \left( - \sum_i \gamma \Delta t (n_i - \beta_t^i )^2 \right) \, ,
\end{equation}
satisfying the completeness condition:
\begin{equation} \label{eq: app mM2}
  \int \mM({\bm \beta_t}) \mM^{\dag}({\bm \beta_t} ) \left( \prod_{i = 1}^N d \beta_t^i \right) = 1 \, .
\end{equation}
According to Born's rule, the probability weight of a quantum trajectory $\mM_{0:t} = \mM({\bm \beta_t}) \mU_{t} \ldots \mM({\bm \beta_{\Delta t}}) \mU_{\Delta t}$ is proportional to ${\rm Tr}( \mM_{0:t} \mM_{0:t}^{\dag})$.
We decompose $\mM_t( {\bm \beta_t} )$ as $\mM_t( {\bm \beta_t} ) = \sqrt{c({\bm \beta_t} )} \tilde{\mM}_t({\bm \beta_t})$ with
\begin{align} 
  \tilde{\mM}_t({\bm \beta_t}) &\equiv \exp \left\{ \gamma \Delta t \sum_i  \left( n_i - \frac{1}{2} \right) (2 \beta_t^i - 1)  
   \right\}, \\
  c({\bm \beta_t} ) &\equiv \left( \frac{2 \gamma \Delta t}{\pi} \right)^{N/2} \exp \left\{ - \frac{\gamma \Delta t}{2} \sum_i  \left[ (2 \beta_t^i - 1)^2 - 1 \right] \right\} \, . 
\end{align}
Consequently, we have 
\begin{equation}
{\rm Tr}( \mM_{0:t} \mM_{0:t}^{\dag}) = \left[  c({\bm \beta}_{\Delta t}) \ldots c({\bm \beta}_{t}) \right] {\rm Tr}( \tilde{\mM}_{0:t} \tilde{\mM}_{0:t}^{\dag}),\quad \tilde{\mM}_{0:t} = \tilde{\mM}({\bm \beta_t}) \mU_{t} \ldots \tilde{\mM}({\bm \beta_{\Delta t}}) \mU_{\Delta t}.
\end{equation}
The first factor, $c({\bm \beta_{\Delta t}}) \cdots c({\bm \beta_{t}})$, is proportional to the probability weight if $\sqrt{\gamma} \Delta t (2 \beta_t^i - 1)$ follows the standard Wiener process. 
Hence, it is proportional to the distribution $p_{F}(\{ z_n\}; t)$ for the quantum trajectory under the forced measurement.

The second factor, ${\rm Tr}(\tilde{\mM}_{0:t} \tilde{\mM}_{0:t}^{\dag})$, is evaluated by considering the single-particle quantum trajectory.
For two generic fermionic Gaussian operators $\mathcal{S} = e^{\sum_{ij} c_i^{\dag} S_{ij} c_j}$ and $\mathcal{Q} = e^{\sum_{ij} c_i^{\dag} Q_{ij} c_j}$, with generic complex matrices $S$ and $Q$, let us introduce $\mathcal{R} = \mathcal{Q} \mathcal{S}$. 
The operator $\mathcal{R}$ is still Gaussian and thus written as $\mathcal{R} = r e^{\sum_{ij} c_i^{\dag} R_{ij} c_j}$, where $r$ and $R$ are a constant and matrix to be determined, respectively.
The matrix $R$ is determined by $e^R = e^S e^Q$ because of $\mathcal{R} c_i^{\dag} \mathcal{R}^{-1} = \sum_j c_j^{\dag} (e^R)_{ji}$ and $\mathcal{Q} \mathcal{S} c_i^{\dag} \mathcal{S}^{-1} \mathcal{Q}^{-1} = \sum_j c_j^{\dag} (e^Q e^S)_{ji}$.
To determine the constant $r$, we observe $\det(\mathcal{S}) = \exp\left[ \Tr\left(\sum_{ij} c_i^{\dag} S_{ij} c_j\right)\right]$. 
In the many-body Hilbert space, we have $\Tr \left( \sum_{ij} c_i^{\dag}S_{ij} c_j \right) = 2^{N-1} \Tr(S)$ with $N$ being the number of fermions,  
and hence $r = \exp \left\{ 2^{N-1} \left[ \Tr(S) + \Tr(Q) - \Tr(R)  \right]  \right\}$.
Meanwhile, we also have $\det(e^R) = \det(e^S) \det(e^Q)$, implying $\Tr(S) + \Tr(Q) = \Tr(R)$.
Consequently, the constant $r$ is determined as $r = 1$. 
The operator $\tilde{\mM}(\bx_t)$ defined earlier is expressed in the form $\tilde{\mM}(\bx_t) = e^{- (1/2)\,\Tr(\epsilon)} e^{\sum_{ij} c_i^{\dag} \epsilon_{ij} c_j}$, where $\epsilon$ is a diagonal matrix with $\epsilon_{ii} = \gamma \Delta t (2 \beta_t^i - 1)$, and its single-particle representation reads $M_t = e^{\epsilon}$.
For $M_{0:t}M_{0:t}^{\dag} = e^{2 P}$ ($M_{0:t} \equiv M_{t} U_{t} \ldots M_{\Delta t} U_{\Delta t} $, $P = P^{\dag}$), we have $\mM_{0:t} \mM_{0:t}^{\dag} = e^{-\Tr(P)} e^{ \sum_{ij} c_i^{\dag} 2P_{ij} c_j} $.
If the eigenvalues of $P$ are denoted by $z_i$'s ($i = 1,2,\ldots, N$), we have $\Tr(M_{0:t}M_{0:t}^{\dag}) = \prod_{n} \left[ 2 \cosh(z_n) \right]$ in the many-body Hilbert space, which is the prefactor in Eq.~(\ref{eq: app p_MD p_RN}).

We also verify this argument by straightforward calculations. Let us substitute 
\begin{equation}
    p_{B}(\{ z_n\}; t) =  e^{-\frac{N}{2} \gamma t} f(\{ z_n \}) p_{F}(\{ z_n\}; t), \quad f(\{ z_n \}) = \left( \prod_{n} \cosh(z_n) \right) 
\end{equation}
into Eq.~(\ref{eq: app FP A0}). 
The left-hand side of Eq.~(\ref{eq: app FP A0}) reads
\begin{equation} 
  \frac{N + 1}{\gamma}  \frac{\partial p_{B}}{\partial t} = 
     \frac{N + 1}{\gamma}  f e^{-\frac{N}{2} t} \left( 
    \frac{\partial p_{F}}{\partial t} 
    - \frac{N}{2} \gamma p_{F}
    \right) \, .
\end{equation}
The right-hand side of Eq.~(\ref{eq: app FP A0}) reads
\begin{align}
    &- \sum_{n=1}^N \frac{\partial  \left[ (\mu_n + \nu_n )p_{B} \right] }{\partial z_n }  + \frac12 \sum_{m,n=1}^N \frac{\partial^2 \left[(1  + \delta_{mn}) p_{B} \right] }{\partial z_n \partial z_m} \nonumber \\
  &\qquad = -f e^{-N/2 \gamma  t}  \left[  \sum_{n}  \frac{\partial (\mu_n  p_{F})   }{\partial z_n} + \frac12 \sum_{m,n} (1 + \delta_{mn})  \frac{\partial^2  p_{F}}{\partial z_n \partial z_m} \right] \nonumber \\
  &\qquad \qquad - f e^{-N/2 t}  p_{F}  \sum_n \sum_{m \neq n} \left[  \frac{1}{2} \tanh(z_n) \tanh(z_m) - \tanh(z_n)  \coth(z_n-z_m) \right] -  N f e^{-N/2 t}  p_{F} .
\end{align}
From the identity 
\begin{equation} 
  \tanh(z_n) \left[  \frac{1}{2}  \tanh(z_m) - \coth(z_n-z_m)  \right] +  \tanh(z_m) \left[  \frac{1}{2}  \tanh(z_n) - \coth(z_m-z_n)  \right] = 1 ,
\end{equation}
the right-hand side of Eq.~(\ref{eq: app FP A0}) is simplified to
\begin{equation} 
    -f e^{-N/2 \gamma  t}  \left[  \sum_{n}  \frac{\partial (\mu_n  p_{F})   }{\partial z_n} + \frac{1}{2} \sum_{m,n} (1 + \delta_{mn})  \frac{\partial^2  p_{F}}{\partial z_n \partial z_m} \right]  -  N(N+1) f e^{-N/2 t}  p_{F} .
\end{equation}
Given the condition that $p_{F}$ is the solution to Eq.~(\ref{eq: app FP A0}) with $\nu_n = 0$, the left-hand and right-hand sides of Eq.~(\ref{eq: app FP A0}) are indeed identical.

\subsection{Monitored dynamics of Majorana fermions}
\label{subsec: MD D}

We consider the dynamics of $2 N$ free Majorana fermions $\psi_i$'s ($\{ \psi_i, \psi_j \} = 2 \delta_{ij}$, $\psi_i = \psi_i^{\dag}$) under Born measurement. 
The unitary dynamics $\mathcal{O}_t =  e^{-\ii \mH_t \Delta t}$ is generated by a time-dependent quadratic Majorana Hamiltonian $\mH_t = \sum_{ij} \psi_i (H_t)_{ij} \psi_j$ ($H_t = H_t^{\dag}$, $H_t = -H_t^{\rm T}$).
The Majorana pairs $\ii \psi_{2j-1} \psi_{2j}$ ($1 \leq j \leq N$) are continuously measured, corresponding to a Kraus operator~\cite{wiseman2009quantum}
\begin{equation} \label{eq: app Mt D}
  \mM_t = e^{ \sum_j \ii \epsilon_j  \psi_{2j-1} \psi_{2j}}, \quad \epsilon_j \equiv \langle  \ii \psi_{2j-1} \psi_{2j}\rangle_t  \gamma d t  + \sqrt{\gamma} d W_t^j  
\end{equation}
with $\gamma$ being the measurement strength and $d W_t^j$ being the standard Wiener process.
The product $\mM_{0:t}= \mM_{t} \mO_{t} \ldots \mM_{\Delta t} \mO_{\Delta t} $ gives a quantum trajectory, and $\rho_t = \mM_{0:t}\mM_{0:t}^{\dag}$.
We introduce a single-particle Kraus operator $M_t$ by $ \mM_t \psi_i \mM_t^{-1}  = \sum_j \psi_j (M_t)_{ji}$, satisfying
\begin{equation} \label{eq: app M_t D}
 M_t = e^{ -2 \sigma_y \otimes {\bm \epsilon} }, \quad {\bm \epsilon} \equiv {\rm diag}(\epsilon_1, \ldots, \epsilon_{N}).
\end{equation}
Here, $M_t$ is written in the basis where $\psi_j$'s are ordered as $ (\psi_1, \psi_3, \ldots, \psi_{2N-1}, \psi_2, \psi_4, \ldots, \psi_{2N})$.  
We also introduce a single-particle unitary operator $O_t$ by  $\mathcal{O}_t  \psi_i \mathcal{O}_t^{-1} = \sum_j \psi_j(O_t)_{ji}$ with $O_t = e^{-4 \ii H_t} \in {\rm SO}(2N)$. 
Owing to Gaussianity, $\rho_t$ is fully encoded in the single-particle quantum trajectory $M_{0:t} \equiv M_{t} O_{t} \ldots M_{\Delta t} O_{\Delta t}$: 
$\rho_t \propto e^{(1/2) \sum_{ij} P_{ij} \psi_i \psi_j}$
with $e^{2 P} \equiv M_{0:t}M_{0:t}^{\dag}$.
The parent Hamiltonian $2P$ is a Hermitian anti-symmetric matrix (see the discussion below), which gives the Majorana two-point correlation: $\ii \langle [ \psi_i, \psi_j]/2 \rangle_t = -\ii \tanh(P)$. 
Due to symmetry, the eigenvalues $2z_n$'s of $P$ appear in $(2z_n, -2z_n)$'s pairs ($z_n \geq 0$), which give the $\alpha$-R\'enyi entropy $S_{\alpha} = \sum_{n} f_{s \alpha} (z_n) $ with $f_{s \alpha}(z) \equiv
 \ln [ {(1 + e^{2z})^{-\alpha}}+{(1 + e^{-2z})^{-\alpha}} ]/(1 - \alpha)$.

We consider symmetry of $M_{0:t}$.
Due to $M_t^{\rm T} = M_t^{-1}$ and $O_t^{\rm T} = O_t^{-1}$, the product $M_{0:t}$ satisfies $M_{0:t}^{\rm T} = M_{0:t}^{-1}$.
The generator $L_{\rm eff}$ of $M_{0:t}$ ($M_{0:t} \equiv e^{L_{\rm eff} t}$) satisfies $L_{\rm eff}^{\rm T} = -L_{\rm eff}$ and hence belongs to non-Hermitian symmetry class D~\cite{kawabata2019}.
Additionally, due to this symmetry, the Hermitian matrix $2P$ also satisfies $P^{\rm T} = P^{-1}$; 
$M_{0:t} M_{0:t}^{\dag}$ is diagonalized as $ M_{0:t} M_{0:t}^{\dag} = Q_t (e^{\sigma_y \otimes 2 {\bm z}}) Q_t^{\rm T} $ with $Q_t \in {\rm SO}(2N)$ and ${\bm z} = {\rm diag}(z_1, \ldots, z_{N})$.

We study the stochastic time evolution of $z_n(t)$'s with the assumption that $O_t$ is distributed uniformly and independently according to the Haar measure on ${\rm SO}(2N)$. 
At time $t + \Delta t$, we have 
\begin{equation}
    M_{0:t+\Delta t} M_{0:t+\Delta t}^{\dag}  = M_{t + \Delta t} O_{t + \Delta t}  P_t e^{\sigma_y \otimes 2 {\bm z}} P_t^{\rm T} O_{t + \Delta t}^{\rm T} M_{t+\Delta t}^{\dag} \, ,
\end{equation}
which shares the same spectrum with $(1 + w) e^{\sigma_z \otimes 2 {\bm z}} (1 + w^{\dag})$.
Here, we define $O \equiv O_{t + \Delta t} P_t$, which should be distributed uniformly in the Haar measure, 
$U \equiv \cfrac{1}{\sqrt{2}} \begin{pmatrix} 
    1 & 1 \\
    \ii & -\ii \\
\end{pmatrix} \otimes 1_{N\times N}$, and $w \equiv  U^{\dag} O^{\rm T} ( M_{t+\Delta t} - 1) O U$, satisfying
\begin{align}
  w &\equiv  U^{\dag} O^{\rm T} ( M_{t+\Delta t} - 1) O U \nonumber \\
  &= 
  -\left(
    \begin{array}{cc}
      \left(A^{\rm T} + \ii B^{\rm T}\right) \be (D + \ii C) +
      \left(D^{\rm T} - \ii C^{\rm T}\right) \be (A - \ii B) & 
        \left(C^{\rm T} + \ii D^{\rm T}\right) \be (B - \ii A)
        -\ii  \left(A^{\rm T} + \ii B^{\rm T}\right) \be (C + \ii D) \\
      \left(B^{\rm T}+\ii A^{\rm T}\right) \be (C-\ii D) 
      +i \left(C^{\rm T}-\ii D^{\rm T}\right) \be (A+\ii B)  & 
     \left(B^{\rm T}+i A^{\rm T}\right) \be (C+i D) +\left(C^{\rm T}-i D^{\rm T}\right) \be (B-i A)  \\
    \end{array}
    \right) \nonumber \\
    &\qquad\qquad + 2 \gamma \Delta t + \mathcal{O}(\Delta t^{3/2}),
\end{align}
with $O \equiv \begin{pmatrix} A & B \\ C & D \end{pmatrix} $ and ${\bm \epsilon}$ being a diagonal matrix, ${\bm \epsilon}_{jj} \equiv \gamma \Delta t \times \langle   \ii \psi_{2j-1} \psi_{2j}\rangle_{t + \Delta t - 0^+} + \sqrt{\gamma} \Delta W_t^j$.
By perturbation theory, we have
\begin{equation} \label{eq: app delta z_n 2}
  2 z_n(t + \Delta t) - 2z_n(t) =  2 w_{nn} + \sum_{m = 1}^{2N} |w_{nm}|^2 +   4 \sum_{ m \neq n, m = 1}^N \frac{|w_{nm}|^2  e^{2 z_m(t)}}{e^{2 z_n(t)} - e^{2 z_m(t)}} 
  + 4 \sum_{  m = N+ 1}^{2N} \frac{|w_{nm}|^2  e^{-2 z_{m-N}(t)}}{e^{2 z_n(t)} - e^{-2 z_{m-N}(t)}} 
  - 2 w_{nn}^2 + \mathcal{O}(\Delta t^{3/2}) \, .
\end{equation}
The correlation function $\langle \ii [\psi_i, \psi_j]/2 \rangle_{t + \Delta t - 0^+}$ is determined by 
$ O (-\ii \sigma_y \otimes \tanh({\bm z}) )O^{\rm T}$
as
\begin{equation} 
  \langle \ii [\psi_i, \psi_j]/2 \rangle_{t + \Delta t - 0^+} = 
  -\ii \begin{pmatrix} 
 \ii B \tanh(\bz) A^{\rm T}- \ii A  \tanh(\bz) B^{\rm T}  &
 \ii B \tanh(\bz) C^{\rm T}- \ii A  \tanh(\bz) D^{\rm T} \\
 \ii D \tanh(\bz) A^{\rm T}- \ii C  \tanh(\bz) B^{\rm T} &
 \ii D \tanh(\bz) C^{\rm T}- \ii C  \tanh(\bz) D^{\rm T} \\
  \end{pmatrix} \, ,
\end{equation}
and
\begin{equation} 
  \langle  \ii \psi_{2j-1} \psi_{2j} \rangle_{t + \Delta t - 0^+} = -2 \sum_m (A_{jm} D_{jm} - B_{jm} C_{jm}) \tanh(z_m).
\end{equation}
We perform the ensemble average over the Haar measure~\cite{brouwer1996} and the Wiener process for each term in Eq.~(\ref{eq: app delta z_n 2}). 
For $n \leq N$, this yields 
\begin{align}
    \langle w_{nn} \rangle_E 
    &= 2 \gamma \Delta t + 4 \sum_{j,m} \left \langle (A_{jn} D_{jn} - B_{jn} C_{jn})  (A_{jm} D_{jm} - B_{jm} C_{jm})    \right \rangle \tanh(z_m)  \gamma \Delta t \nonumber \\
    &= \frac{4 }{2 N - 1} \tanh(z_n) \gamma \Delta t + 2 \gamma \Delta t \, .
\end{align}
Additionally, for $n, m \leq N$, we have $ \langle w_{nn} w_{mm} \rangle_E = 4 \gamma \Delta t  \delta_{mn}/\left( 2N-1 \right)$; 
for $|n - m| = N$, $w_{nm} = 0$; for $|n - m| \neq N$, $\langle |w_{mn}|^2  \rangle_E = 4 \gamma \Delta t/\left( 2N-1 \right)$. 
Substituting these results into Eq.~(\ref{eq: app delta z_n 2}), we have 
\begin{align} \label{eq: app EoM z D}
  \langle \Delta z_n(t) \rangle_E 
   &=   \frac{4 \left(  \mu_n  + \nu_n \right) }{2N - 1} \gamma \Delta t  ,\\  
  \langle \Delta z_n(t) \Delta z_m(t) \rangle_E  &= \frac{4 \delta_{mn}}{2 N-1} \gamma \Delta t  , \label{eq: app EoM z2 D}
\end{align}
with
\begin{equation} 
  \mu_n \equiv \sum_{m \neq n} \left( \coth(z_n - z_m) + \coth(z_n + z_m) \right) , \quad \nu_n \equiv \tanh(z_n) \,.
\end{equation}
The resulting Fokker-Planck equation is 
\begin{equation} \label{eq: app DMPK D}
\begin{aligned} 
     \frac{2N-1}{4 \gamma} \frac{\partial p}{\partial t} & = - \sum_{n=1}^{N} \frac{\partial (\mu_n + \nu_n) p }{\partial z_n} + \frac{1}{2} \sum_{n=1}^{N}     \frac{\partial^2 p }{\partial z_n^2}  .
\end{aligned}
\end{equation}

If the Born measurement is replaced by the forced measurement, the Kraus operator $\mM_t$ still takes the same form as Eq.~(\ref{eq: app Mt D}), but with  $\epsilon_j = \sqrt{\gamma}dW_t^j$.  
By a similar method, we find that the Fokker-Planck equation for forced measurement is Eq.~(\ref{eq: app DMPK D}) with $\nu_n = 0$. 
Equation~(\ref{eq: app DMPK D}) with $\nu_n = 0$ also arises in the quantum transport, and we find its exact solution with the initial condition $p_{F}(\{z_i\};t = 0) = \delta(z_1) \delta(z_2) \ldots \delta(z_N) $~\cite{brouwer2000}:
\begin{equation} 
  p_{F}(\{z_i\};t) =   \mathcal{N}(t) \left( \prod_{j<k} (z_j^2 - z_k^2) (\sinh^2 z_j - \sinh^2 z_k) \right) \prod_j e^{-  (2N -1 ) z_j^2/(8 \gamma t)  }.
\end{equation}
with a normalization constant $\mathcal{N}(t)$. 
Following the same argument in Sec.~\ref{subsec: exact solution} and performing straightforward calculations, we find that 
\begin{equation} 
  p_{B}(\{z_i\};t) =   e^{- N\gamma t} \prod_i \cosh(z_i) p_{F}(\{z_i\};t) 
\end{equation}
is the exact solution to the Fokker-Planck equation for the Born measurement with the same initial condition.

We investigate the Lyapunov exponents $\eta_n = \lim_{t \to \infty}\langle z_n \rangle_E/t$ of the quantum trajectory $M_{0:t}$ by using the mean-field solutions.
For the forced measurement, given that $\nu_n = 0$ and non-negative $z_n$'s are well separated, we have from Eq.~(\ref{eq: app EoM z D})
\begin{equation}
    \braket{z_n}_E = \frac{8 \left( n-1 \right)}{2N-1} \gamma t.
\end{equation}
Thus, a Lyapunov zero eigenvalue $\eta_1$ always exists without the even-odd effect, implying the divergent purification time.
This contrasts with complex fermions, but is similar to disordered superconductors in class DIII~\cite{brouwer2000}.
For the Born measurement, due to the presence of $\nu_n = \tanh(z_n) \simeq {\rm sign}(z_n)$, we instead have
\begin{equation}
    \braket{z_n}_E = \frac{4 \left( 2n-1 \right)}{2N-1} \gamma t \, ,
\end{equation}
which is non-zero for any $n$ and $N$.
This implies that $\langle S_{\alpha} \rangle_E$ always decays exponentially with time, similar to complex fermions under Born measurement.

\subsection{Monitored dynamics with enriched symmetry}
\label{subsec: MD AI and BDI}
For the monitored dynamics of $N$ complex fermions, we consider the Hamiltonian $H_t$ with particle-hole symmetry (i.e., $H_t^{\rm T} = - H_t$). 
Note that we should not confuse this symmetry with particle-hole symmetry in the Majorana Hamiltonian, which is just a consequence of the Majorana basis.
The single-particle representation of the unitary operator reads $U_t = e^{-\ii H \Delta t} \in {\rm SO}(N)$. 
Since the single-particle Kraus operator $M_t$ in Eq.~(\ref{eq: app M_t}) is also real, $M_{0:t}^* = M_{0:t}$, its generator is also real, $L_{\rm eff}^* = L_{\rm eff}$, resulting in non-Hermitian symmetry class AI~\cite{kawabata2019}. 
For the Born measurement, following the same procedure in Sec.~\ref{subsec: MD A}, we find the Fokker-Planck equation for the distribution $p(\{z_n\};t)$,
\begin{equation} \label{eq: app FP AI}
  \frac{N + 2}{\gamma} \frac{\partial p}{\partial t} 
  = - \sum_{n=1}^N \frac{\partial  \left[ (\mu_n + \nu_n )p \right] }{\partial z_n }  + \frac{1}{2} \sum_{m,n=1}^N \frac{\partial^2 \left[(1  + 2 \delta_{mn}) p \right] }{\partial z_n \partial z_m} \, 
\end{equation}
with
\begin{equation} \label{eq: mu nu AI}
\mu_n \equiv \sum_{m \neq n} \coth(z_n-z_m), \quad \nu_n \equiv  \sum_m  (1 + 2 \delta_{nm}) \tanh(z_m(t)) .
\end{equation}
For the forced measurement, the resulting Fokker-Planck equation takes the same form as Eq.~(\ref{eq: app FP AI}) but with $\nu_n = 0$.
After changing the variables by
\begin{equation} 
  y_n \equiv \sum_{m}A_{nm} z_m~\text{ with }~A_{nm} \equiv -\frac{1}{N} + 
  \frac{\sqrt{2}}{\sqrt{N+2} N} +\delta_{nm}, \quad s \equiv  \frac{2 \gamma t}{N+2} \, , 
\end{equation}
Eq.~(\ref{eq: app FP AI}) with $\nu_n = 0$ also appears in the quantum transport~\cite{brouwer1998}.

For the monitored dynamics of $2N$ Majorana fermions, we consider $H_t$ with a block-diagonalized structure: $\sigma_z H_t \sigma_z = H_t$ and $H_t^{\rm T} = - H_t$, which leads to
$\sigma_z U_t \sigma_z = U_t$
besides $U_t \in {\rm SO}(2N)$.
Meanwhile, the single-particle Kraus operator $M_t$ in Eq.~(\ref{eq: app M_t D}) satisfies $\sigma_z M_t^{*} \sigma_z = M_t$ and $M_t^{\rm T} = M_t^{-1}$. 
Consequently, we have $\sigma_z M_{0:t}^{*} \sigma_z = M_{0:t}$ and $M_{0:t}^{\rm T} = M_{0:t}^{-1}$, as well as, $\sigma_z L_{\rm eff}^* \sigma_z = L_{\rm eff}$ and $L_{\rm eff}^{T} = L_{\rm eff}^{-1}$, resulting in class BDI~\cite{kawabata2019}. 
For the Born measurement, following the same procedure in Sec.~\ref{subsec: MD D}, we find the Fokker-Planck equation for the distribution $p(\{z_n\};t)$,
\begin{equation} \label{eq: app FP BDI}
\begin{aligned} 
     \frac{N}{2 \gamma} \frac{\partial p}{\partial t} & = - \sum_{n=1}^{N} \frac{\partial (\mu_n + \nu_n) p }{\partial z_n} + \frac{1}{2} \sum_{n=1}^{N}     \frac{\partial^2 p }{\partial z_n^2}  .
\end{aligned}
\end{equation}
with
\begin{equation} \label{eq: mu nu BDO}
  \mu_n \equiv \sum_{m \neq n} \left( \coth(z_n - z_m) + \coth(z_n + z_m) \right) , \quad  \nu_n \equiv 2 \tanh(z_n) \,.
\end{equation}
For the forced measurement, the resulting Fokker-Planck equation takes the same form as Eq.~(\ref{eq: app FP BDI}) but with $\nu_n = 0$, which also appears in the quantum transport~\cite{brouwer2000}.

\section{
Purification dynamics}
\label{sec: UEF}
\subsection{Mean-field solutions}
\label{subsec: MF maximum}
We use the distribution $p_B(\{ z_n\};t)$ in Eqs.~(\ref{eq: app p_RN}) and (\ref{eq: app p_MD p_RN}) for complex fermions under Born measurement as an example to demonstrate that, in the long-time limit $t \rightarrow \infty$, analyzing $z_n$'s that maximize the distribution $p(\{ z_n\};t)$ is equivalent to finding mean-field solutions to the Fokker-Planck equations.
The maxima of $p_B(\{ z_n\};t)$ correspond to the minima of $W(\{ z_n \}) \equiv -\ln(p_{B}(\{z_n\};t) )$:
\begin{equation} 
  W(\{ z_n \}) =  -\sum_{n} \ln\left( \cosh(z_n) \right) -  \sum_{n<m} \ln \left[ (z_n - z_m) \sinh(z_n - z_m) \right]  + \frac{N+1}{2 \gamma t} \sum_{n,m} z_n \left( -\frac{1}{N+1} + \delta_{nm} \right) z_m \, .
\end{equation}
A minimum of $W(\{ z_n \};t)$ requires ${\partial W}/{\partial z_n} = 0$ ($n = 1,2, \ldots, N$) with  
\begin{equation}  \label{eq: app min W}
  \frac{\partial W}{\partial z_n}
=  -\tanh(z_n) - \sum_{m \neq n} \left[ \frac{1}{z_n - z_m} +  \coth(z_n - z_m) \right]+ \frac{N+1}{\gamma t} \sum_{m} \left( -\frac{1}{N+1} + \delta_{nm} \right)z_m .
\end{equation}
Summing over all $n = 1, 2, \ldots, N$ in Eq.~(\ref{eq: app min W}) gives
\begin{equation} \label{eq: app sum z}
 \sum_{n = 1}^{N} z_n = \gamma t \sum_{n = 1}^N \tanh(z_n) .
\end{equation}
Substituting Eq.~(\ref{eq: app sum z}) into Eq.~(\ref{eq: app min W}), we obtain
\begin{equation} \label{eq: app z_n}
  z_n = \frac{\gamma t }{N + 1}  \sum_{m \neq n} \left[ \frac{1}{z_n - z_m} + \coth(z_n - z_m) + \tanh(z_m) \right] + \frac{2\gamma t}{N + 1} \tanh(z_n) \, .
\end{equation}
We note that the mean-field solution [\MF]
\begin{equation} \label{eq: app z_n2}
  \langle z_n  \rangle_E = \frac{2 (n -l)- 1 + {\rm sign}(n-l-0^+) }{N + 1} \gamma t
\end{equation}
in the long-time limit, satisfies $1/(\langle z_n \rangle_E - \langle z_m \rangle_E) \propto 1/t \ll 1$, $\coth(\langle z_n \rangle_E - \langle z_m \rangle_E) = {\rm sign}(n-m) $ ($n \neq m$), and $\tanh(\langle z_m \rangle_E) = {\rm sign}(m - l - 0^+)$.
Thus, these $\langle z_n \rangle_E$'s satisfy Eq.~(\ref{eq: app z_n}) and hence represent a local minimum of $W(\{ z_n\})$.

Additionally, for $N \gg 1$, in the time regime $1 \ll \gamma t \ll N$, the mean filed solution~(\ref{eq: app z_n2}) still approximately represents a local maximum of $p_B(\{ z_n\};t)$.  
In such a time regime, for $ |n - m| \gg N/\gamma t $, we have $1/(\langle z_n \rangle_E - \langle z_m \rangle_E) \simeq \left( N+1 \right)/2 (n-m) \gamma t \ll 1$ and $\coth(\langle z_n \rangle_E - \langle z_m \rangle_E) \simeq {\rm sign}(n-m) $; 
for $ |m - l| \gg N/\gamma t$, we have $\tanh(\langle z_m \rangle_E) = {\rm sign}(m - l)$. 
Due to $\gamma t \gg 1$, for given $m$ (or $l$), most of $n \in [1,N]$ satisfy $|n - m| \gg N/\gamma t$ (or $|n - l| \gg N/\gamma t$), resulting in $\langle z_n \rangle_E$'s approximately satisfying Eq.~(\ref{eq: app z_n}).

\subsection{Weight of mean-field solutions for Born measurement}
\label{subsec: MF weight}
For complex fermions under Born measurement, we calculate the weight of the $l$th ($l = 0,1,\ldots, N$) mean-field solutions [Eq.~(\ref{eq: app z_n2})] in the long-time limit.
This is achieved by calculating the ensemble average of the probability of having $N - l$ particles in $\rho_t$, denoted by $\langle {\rm Pr}(n_{\rm tot} = N- l) \rangle_E$, according to Born's rule.
As discussed in Sec.~\ref{subsec: exact solution}, the continuous measurement on $n_i$'s corresponds to a complete set of Kraus operators $\mM(\beta_t)$ [Eqs.~(\ref{eq: app mM}) and (\ref{eq: app mM2})] with the measurement outcome $\bx_t$. 
If the initial density matrix is $\rho_0 =  \mathbb{1}/\Tr(\mathbb{1})$ and the measurement outcomes are $\{ \bx \} = \{ \bx_{\Delta t }, \bx_{2 \Delta t}, \ldots, \bx_{t}\}$, the un-normalized density matrix at $t$ is $\rho_{\{\bx \}} = \mM(\{\bx\}) \mM^{\dag}(\{\bx\})/\Tr(1) $ with $ \mM(\{\bx\}) = \mM(\bx_t) \mU_t \ldots  \mM(\bx_{\Delta t}) \mU_{\Delta t} $.
According to Born's rule, the probability of $\rho_t$ being $\rho_{\{\bx \}}$ is proportional to $\Tr ( \rho_{\{\bx \}})$. 
Additionally, in $\rho_{\{\bx \}}$, the probability of having $N - l$ particles, ${\rm Pr}(n_{\rm tot} = N - l)$, is $  \Tr ( \rho_{\{\bx \}} \mathbb{1}_l)/ \Tr ( \rho_{\{\bx \}})$, where $\mathbb{1}_l$ is the projection operator to the $(N-l)$-particle subspace. 
Performing the ensemble average over ${\rm Pr}(n_{\rm tot} = N - l)$, we have 
\begin{equation} 
 \langle {\rm Pr}(n_{\rm tot} = N - l) \rangle_E= \int   \frac{\Tr ( \rho_{\{\bx \}} \mathbb{1}_p)}{\Tr ( \rho_{\{\bx \}})} \times  \Tr ( \rho_{\{\bx \}})\,d \beta_{\Delta t} \cdots d \beta_{t}.
\end{equation} 
Since both $\mU_t$ and $\mM(\bx_t)$ commute with $\mathbb{1}_p$, we have
\begin{equation} \label{eq: app trace rho}
     \Tr ( \rho_{\{\bx \}} \mathbb{1}_p) =  \frac{\Tr \left[\mathbb{1}_p 
     \mU_{\Delta t}^{\dag} \mM(\beta_{\Delta t})^{\dag} 
       \ldots \mU_t^{\dag} K(\beta_t)^{\dag} K(\beta_t) \mU_t \ldots \mM(\beta_{\Delta t}) \mU_{\Delta t} \right]}{\Tr(\mathbb{1})} .
\end{equation}
As a result of Eq.~(\ref{eq: app trace rho}) and the completeness of the Kraus operators [Eq.~(\ref{eq: app mM2})], we further have
\begin{equation}
     \langle {\rm Pr}(n_{\rm tot} = N - l) \rangle_E =  \int   \Tr ( \rho_{\{\bx \}} \mathbb{1}_p)   d \beta_{\Delta t} \cdots d \beta_{t} = \frac{\Tr(\mathbb{1}_p)}{\Tr(\mathbb{1})}. 
\end{equation}

Next, we evaluate $\langle {\rm Pr}(n_{\rm tot} = N - l) \rangle_E$ using the mean-field solutions. 
The total particle number is $\langle n_{\rm tot} \rangle_t = \Tr(\rho n_{\rm tot})/\Tr(\rho) = \sum_{i} (\tanh z_i(t) + 1)/2$. 
Additionally, all $|\langle z_i \rangle|$'s satisfy $|\langle z_i \rangle| \gg 1$.
Thus, for $\rho_t$ around the $l$th mean-field solution, ${\rm Pr}(n_{\rm tot} = m)$ is almost $1$ ($0$) for $m = N -l$ ($m \neq N -l$). 
By averaging ${\rm Pr}(n_{\rm tot} = N - l)$ over all the $N+1$ mean-field solutions, the weight of the $l$th mean-field solution equals $\langle {\rm Pr}(n_{\rm tot} = N - l) \rangle_E = \Tr (\mathbb{1}_l)/\Tr(\mathbb{1}) = C_N^l/2^N$.

\subsection{Universal entropy fluctuations in the short-time regime}
\label{subsec: Var S1}

For complex fermions under Born measurement without any symmetry (Sec.~\ref{subsec: MD A}), we calculate ${\rm Var}(S_{\alpha})  \equiv \langle S_{\alpha}^2 \rangle_E - \langle S_{\alpha} \rangle_E^2  $ in the short-$t$ and large-$N$ regime ($1 \ll \gamma t \ll N$), using the exact solution to the Fokker-Planck equation. 
For the forced measurement in the same regime, ${\rm Var}(S_{\alpha})$ can be evaluated by the same method, which is identical to that for the Born measurement.

We begin with expanding $W \equiv \ln[p_B(\{z_n\};t)]$ by $y_n \equiv z_n  - \langle z_n \rangle_E $.  
Although there exist $N+1$ different mean-field $\langle z_n \rangle_E$'s labeled by $l = 0,1,\ldots, N$, the weight of the $l$th solution is $C_N^l/2^N$ and hence vanishes for $N \gg 1$ and $(l-N/2)/\sqrt{N} \gg 1$, implying that considering the mean-field solution with $l = N/2$ suffices.
To the lowest order, the distribution $p_B(\{z_n\};t)$ is
\begin{equation} 
  p_B(\{z_n\};t) \propto \exp\left[ - \frac{1}{2} \sum_{n,m}\left(  \left. \frac{\partial^2 W}{\partial z_n \partial z_m} \right \rvert_{\{ z_j \} = \{ \langle  z_j\rangle_E \}}  \right)y_n y_m \right] \, ,
\end{equation}
taking a Gaussian form.
Consequently, we have $\langle y_n y_m \rangle_E = (\omega^{-1})_{nm}$ with 
\begin{align} \label{eq: app W_nm}
 \omega_{nm} 
 &\equiv \left. \frac{\partial^2 W}{\partial z_n \partial z_m} \right \rvert_{\{ z_j \} = \{ \langle  z_j\rangle_E \}} \nonumber \\
&=  -\sech^2\left(\langle z_n \rangle_E \right) \delta_{nm} - 
\left[ \left(\langle z_n \rangle_E - \langle z_m \rangle_E \right)^{-2} + 
\csch^2\left(\langle z_n \rangle_E - \langle z_m \rangle_E \right) \right] (1 - \delta_{nm}) + \frac{1}{\gamma t}\left[ (N+1)\delta_{nm} -1\right] \, .
\end{align}
We apply the Fourier transformation: 
$Y_k \equiv N^{-1/2} \sum_n e^{-\ii k n} y_n$ ($k = 2 \pi m/N$; $m = 0,1,\ldots,N-1$). 
The Fourier transformation of the matrix $\omega$ is given as
\begin{align} \label{eq: app w_kp}
   \tilde{\omega}_{k,p} & \equiv \frac{1}{N}\sum_{n, m} e^{\ii k n} w_{nm} e^{-\ii p m} \nonumber \\
   & = \frac{1}{N} \sum_n e^{\ii n (k-p)} \left( -\sech^2\left(\langle z_n \rangle_E \right) \right) -  \sum_{(n - m) \neq 0} e^{\ii k (n - m)}  \left[ \left(\langle z_n \rangle_E - \langle z_m \rangle_E \right)^{-2} + 
   \csch^2\left(\langle z_n \rangle_E - \langle z_m \rangle_E \right) \right] \delta_{k,p} \nonumber \\
   &\qquad \qquad - \frac{N}{\gamma t} \delta_{k,0} \delta_{p,0} + \frac{N + 1}{\gamma t} \delta_{k,p} .
 \end{align}
We define $a \equiv (N + 1)/2 \gamma t$ and evaluate each term in $ \tilde{\omega}_{k,p}$ by replacing the sum with an integral. 
Among these terms, 
\begin{align} 
   &\sum_{(n - m) \neq 0} e^{\ii k (n - m)}  \left[ \left(\langle z_n \rangle_E - \langle z_m \rangle_E \right)^{-2} + 
   \csch^2\left(\langle z_n \rangle_E - \langle z_m \rangle_E \right) \right] \nonumber \\ 
   &\qquad =  \sum_{(n - m) \neq 0} e^{\ii k (n - m)}  \left[ \left( \frac{2(n-m) \gamma t}{N+1}\right)^{-2} + 
   \csch^2\left( \frac{2(n-m) \gamma t}{N+1}\right) \right] \nonumber \\
   &\qquad \simeq \int_{-\infty}^{\infty} dx e^{\ii k x} \left[ \left( \frac{x}{a} \right)^{-2} + \csch^2\left( \frac{x}{a} \right) \right] \nonumber \\
   &\qquad =    -\frac{2 \pi   a^2 |k|}{1 - e^{-\pi  a |k|}} \, ,
  \end{align} 
is the leading term and proportional to $(N+1)^2$.
The other terms in $ \tilde{\omega}_{k,p}$ are of order $\mathcal{O}(N^1)$ or $\mathcal{O}(N^0)$.
Thus, in the leading order, we have
\begin{equation} 
 \tilde{
 \omega}_{k,p} \simeq \frac{2 \pi   a^2 |k|}{1 - e^{-\pi  a |k|}} \delta_{k,p}, 
\end{equation} 
and $\langle Y_k Y_p^* \rangle = \delta_{k,p} ( \tilde{
\omega}_{k,k})^{-1} $. 
Consequently, ${\rm Var}(S_{\alpha})$ is 
\begin{align} 
     {\rm Var}(S_{\alpha}) & = \sum_{m,n} f_{s \alpha}^{\prime}(\langle z_m \rangle)
     f_{s \alpha}^{\prime}(\langle z_n \rangle) \langle y_m y_n \rangle \nonumber \\
      &= \sum_{k} \frac{1}{N} \langle  |Y_k|^2 \rangle 
     \sum_m  f_{s \alpha}^{\prime}(\langle z_m \rangle) e^{\ii k m}
     \sum_n  f_{s \alpha}^{\prime}(\langle z_n \rangle) e^{-\ii k n} \nonumber \\
     & \simeq  \sum_{k} \frac{1}{N} \left( \tf_{s \alpha}(ak) a^2 k\right)^2    \frac{1 - e^{-\pi  a k}}{2 \pi a^2 |k|} \nonumber \\
      &=  \int_{-\infty}^{\infty} dq  \frac{|q| (1 - e^{- \pi |q| })}{4 \pi^2} \tf_{s \alpha}(q )^2 \,
  \end{align}
with $\tf_{s \alpha} (k) \equiv \int_{-\infty}^{\infty} f_{s \alpha}(z) e^{-\ii k z} dz$.
For $S_2$, we have 
\begin{equation}
    \tf_{s2} (k) = \frac{\pi \tanh \left({\pi  k}/{8}\right)}{k \cosh \left({\pi  k}/{4}\right)}, \quad {\rm Var}(S_2) = 2 \sigma_2^2 \equiv 10 \ln 2 - 6 \ln \pi =  0.06309 \ldots. 
\end{equation}
For $S_{\infty}$, we have
\begin{equation}
    \tf_{s\infty} (k) = \frac{2}{k^2} \left( 1 - \frac{\pi k}{2 \sinh \left( \pi k/2 \right)} \right), \quad {\rm Var}(S_{\infty}) = 0.04841 \ldots. 
\end{equation}
For some other $S_{\alpha}$'s, we have
\begin{equation} \label{eq: app varS}
 {\rm Var}(S_{\alpha}) = 2 \sigma_{\alpha}^2 = \begin{cases} 
  0.06180\ldots \quad & 
  (\alpha = 1; \text{von Neumann entropy}), \\
  10 \ln 2 -6 \ln \pi =  0.06309 \ldots \quad & 
  (\alpha = 2), \\
  0.06163\ldots \quad & 
  (\alpha = 3), \\
  0.06011 \ldots \quad & 
  (\alpha = 4), \\
   0.04841 \ldots  \quad & 
   (\alpha = \infty).
  \end{cases}
\end{equation}

\subsection{Linear approximation of the Fokker-Planck equation}

We calculate ${\rm Var}(S_{\alpha})$ in the short-time regime using a complementary method to that described in Sec.\ref{subsec: Var S1}.
We evaluate $\langle y_n y_m \rangle_E$ by the linear approximation of the Fokker-Planck equation~\cite{chalker1993}, which is equivalent to the linear approximation of the time evolution of $\Delta z_n$.  
This method is useful when an exact solution to the Fokker-Planck equation is unavailable.
To demonstrate this method, we use complex fermions under Born measurement as an example. 
It can be easily generalized to other monitored dynamics with enriched symmetry.

We expand Eqs.~(\ref{eq: app EoM z}) and (\ref{eq: app EoM z2}) by $y_n \equiv z_n - \langle z_n \rangle_E $.  
Retaining only the linear order in $y_n$, we have  
\begin{align} \label{eq: app linear dy}
   \langle \Delta y_n \rangle_E  & =  \frac{\gamma \Delta t }{N+1} \left[
  \sech^2(\langle z_n \rangle) y_n +   
  \sum_m  \sech^2(\langle z_m \rangle) y_m -  \sum_{m \neq n} \csch^2(\langle z_n - z_m \rangle) (y_n - y_m)  \right],  \\
   \langle \Delta y_n(t) \Delta y_m(t) \rangle_E & =  \frac{1  + \delta_{mn}}{N + 1}\gamma \Delta t \, .
\end{align}
The term $\gamma \Delta t\,\sech^{2}(\langle z_n \rangle_E) y_n/ (N+1)$ on the right-hand side of Eq.~(\ref{eq: app linear dy}) can be omitted due to the presence of $2N \gg 1$ other terms. 
With $a \equiv (N + 1)/2 \gamma t$, $Y_k \equiv N^{-1/2} \sum_n e^{-\ii k n} y_n$ for $k \neq 0$ satisfies
\begin{align} 
   \langle \Delta Y_k \rangle_E 
   &\simeq  -\frac{\gamma \Delta t }{N+1} Y_k \sum_{(n - m) \neq 0 } \left( 1 - e^{-\ii k (n-m)} \right) \csch^2\left(\frac{2(n-m) \gamma t}{N+1} \right)  \nonumber \\
   &\simeq   -\frac{\gamma \Delta t }{N+1} Y_k \sum_{(n - m) \neq 0 } 
   \int  2  \sin^2\left( \frac{kx}{2} \right)\csch^2 \left( \frac{x}{a} \right) dx  \nonumber \\  
   &= \frac{ \Delta t }{2 t} Y_k \left[ 2   - \pi a k \coth\left(  \frac{\pi a k}{2} \right)
   \right]  \, ,
\end{align}
and 
\begin{equation} 
  \langle \Delta Y_k  \Delta Y_{-k} \rangle_E = \frac{\gamma \Delta t}{N + 1}\, .
\end{equation}
Consequently, we have
\begin{equation} \label{eq: app dYk}
  \frac{d  \langle |Y_k|^2 \rangle_E}{dt}   = \frac{ 1 }{ t}\left[ 2   - \pi a k \coth\left(  \frac{\pi ak}{2} \right)
  \right] \langle |Y_k|^2 \rangle_E + \frac{\gamma }{N+1} \, .
\end{equation}
With the initial condition $\langle |Y_k|^2 (t = 0)\rangle_E = 0 $, the solution to Eq.~(\ref{eq: app dYk}) is 
\begin{equation} 
  \langle |Y_k|^2 \rangle = \frac{1 - e^{-\pi a |k|}}{2 \pi a^2 |k|} \, .
\end{equation}
Then, we can follow the same procedure as in Sec.~\ref{subsec: Var S1} to evaluate ${\rm Var}(S_{\alpha})$. 

Applying this method to the monitored dynamics with $L_{\rm eff}$ in class D (Sec.~\ref{subsec: MD D}) and $L_{\rm eff}$ in classes AI and BDI (Sec.~\ref{subsec: MD AI and BDI}), we find
\begin{equation} \label{eq: app varS2}
{\rm Var}(S_{\alpha}) = \begin{cases} 
 \sigma_{\alpha}^2 & \text{(class D)}, \\ 
 4 \sigma_{\alpha}^2 & \text{(class AI)}, \\ 
  2\sigma_{\alpha}^2 & \text{(class BDI)}, 
  \end{cases}
\end{equation}
with $\sigma_{\alpha}^2$ given by Eq.~(\ref{eq: app varS}).

\label{subsec: Var S2}

\section{Numerical simulation}
\label{sec: numeric}
\subsection{Numerical details for weak measurement}
\label{subsec: weak m}
In the numerical simulation, we consider a discrete version of the monitored dynamics of complex fermions. 
A set of complete Kraus operators $\mM_{i; \pm}$ for weak measurement on the particle number $n_i = c_i^{\dag} c_i$ is   
\begin{equation} 
 \mM_{i; \pm} = \frac{e^{\pm \sqrt{\gamma \Delta t}  (n_i - \frac12)}}{\sqrt{2 \cosh(\sqrt{\gamma \Delta t})}} \, ,
\end{equation}
which is complete, i.e., $ \sum_{s = \pm} \mM_{i;s}\mM_{i;s}^{\dag}  = 1$.
According to Born's rule, for a density matrix $\rho_t$, the probability $p_{i;\pm}$ of the measurement outcome $\pm$ is $\Tr(\mM_{i; \pm } \rho_t \mM_{i; \pm})/\Tr(\rho_t)$. 
Given that $\langle  n_i \rangle_t \equiv \Tr( n_i \rho_t)/\Tr(\rho_t)$, we have 
\begin{equation} \label{eq: app p cosh}
 p_{i;+} = \frac{e^{ \sqrt{ \gamma \Delta t}} \langle n_i \rangle_t + e^{ -\sqrt{ \gamma \Delta t}}(1 -\langle n_i \rangle_t)}{2 \cosh(\sqrt{\gamma \Delta t})}, \quad  
 p_{i;-} = \frac{e^{ -\sqrt{ \gamma \Delta t}}\langle n_i \rangle_t + e^{ \sqrt{ \gamma \Delta t}}(1 -\langle n_i \rangle_t)}{2 \cosh(\sqrt{\gamma \Delta t})} \, .
\end{equation}
In our setup, at time $t$, all the sites $n_i$ ($1 \leq i \leq N$) are measured.
The Kraus operator $\mM_t  = \prod_{i=1}^N \mM_{i; s_i} $ and the probability of each result $s_i = \pm$ is given in Eq.~(\ref{eq: app p cosh}), independently. 
When $\gamma$ is fixed and $\Delta t \rightarrow 0$ (i.e., measurement frequency goes to $\infty$), this discrete scheme reduces to the continuous formalism discussed earlier. 
However, for numerical efficiency, we simulate the dynamics discretely and choose $\Delta t = 1$.

In the numerical simulation, we need to calculate the single-particle Kraus operator $M_{0:t} = M_{t} U_{t} \ldots M_{\Delta t} U_{\Delta t}$ and evaluate its singular values $e^{z_i}$'s, where $M_t$ is the single-particle representation of $\mM_t$.
Calculating it directly will lead to a large round error. 
Instead, at each time step, we perform QR decomposition. 
Let us introduce QR decomposition by $M_{0:\Delta t} = Q_{0:\Delta t} R_{0: \Delta t}$, where $ Q_{0: \Delta t} $ is a unitary matrix, and $R_{0: \Delta t}$ is an upper-triangular matrix. 
At the next time step $2 \Delta t$, the quantum trajectory is updated as $M_{0:2 \Delta t} = M_{2 \Delta t} U_{2 \Delta t}  M_{0:\Delta t}$. 
QR decomposition $ Q_{0:2\Delta t} R_{0: 2\Delta t} $ of $M_{0:2 \Delta t}$ is determined as follows.
The matrix $ Q_{0:2 \Delta t}$ is obtained by performing QR decomposition as $M_{2 \Delta t} U_{2 \Delta t} Q_{0:\Delta t} \equiv Q_{0:2 \Delta t} R_{2 \Delta t} $. The matrix $R_{0:2\Delta t}$ is updated as $R_{0: 2 \Delta t} = R_{2 \Delta t} R_{0:\Delta t}$.
The subsequent $M_{0:2\Delta t}$, $\ldots$, $M_{0:t - \Delta t}$ can be calculated similarly.
The resulting $M_{0:t - \Delta t} = Q_{0: t - \Delta t} R_{0:t - \Delta t}$ shares the same singular values with $R_{0:t - \Delta t}$.
Notably, the distribution of $M_{t}$ is determined by the quantum trajectory $U_t M_{t -\Delta t}$, according to the correlation function 
\begin{equation}
C_{ij}(t - 0^+) \equiv \langle  c_i^{\dag} c_j \rangle_{t - 0^+} =\left[ U_t \left( 1 +  \left( M_{0:t-\Delta t}M_{0:t-\Delta t}^{\dag} \right)^{-1}  \right)^{-1} U_t^{\dag} \right]_{ji}
\end{equation}
and Born's rule [Eq.~(\ref{eq: app p cosh})]. 
Performing the singular-value decomposition, $R_{0:t-\Delta t} = A_{t -\Delta t} \Lambda_{t - \Delta t} B_{t-\Delta t}$ [$\Lambda_t = {\rm diag}(e^{z_1},e^{z_2},\ldots, e^{z_N})$], we have $C^{\rm T} = U_t Q_{0:t-\Delta t} A_{t - \Delta t} (1 + \Lambda_{t - \Delta t}^{-2})^{-1}A_{t - \Delta t}^{\dag} Q_{0:t-\Delta t}^{\dag}U_t^{\dag}$.

We also consider a discrete version of the monitored dynamics of Majorana fermions.
A set of complete Kraus operators $\mM_{i; \pm}$ for weak measurement on the Majorana pair $ \ii \psi_{2i - 1} \psi_{2i}$ is 
\begin{equation} 
 \mM_{i; \pm} = \frac{e^{ \ii \sqrt{\gamma \Delta t} \psi_{2i - 1} \psi_{2i} }}{\sqrt{2 \cosh(2 \sqrt{\gamma \Delta t})}} .
\end{equation}
The probability $p_{i;\pm}$ of the measurement result $\pm$ is
\begin{equation} \label{eq: app p cosh2}
 p_{i;+} = \frac{ \cosh( 2\sqrt{ \gamma \Delta t}) +  \langle \ii \gamma_{2i -1} \gamma_{2i} \rangle \sinh(2 \sqrt{\gamma \Delta t})}{2 \cosh(2 \sqrt{\gamma \Delta t})}, \quad  
  p_{i;-} = \frac{ \cosh( 2 \sqrt{\gamma \Delta t}) - \langle \ii \gamma_{2i -1} \gamma_{2i} \rangle \sinh(2 \sqrt{\gamma \Delta t})}{2 \cosh( 2 \sqrt{\gamma \Delta t})} \, .
\end{equation}
We also choose $\Delta t= 1$ and use the QR decomposition method to calculate the single-particle Kraus operator $M_{0: t}$ for the monitored dynamics of Majorana fermions.

In the simulation of forced measurement, the Born probability [Eqs.~(\ref{eq: app p cosh}) and (\ref{eq: app p cosh2})] is replaced by the prior probability $p_+ = p_- = 1/2$, while the rest of the procedures remains the same as in the Born measurement.

\subsection{Numerical details for projective measurement}
\label{subsec: projective m}
We numerically simulate monitored dynamics with projective measurement. 
For complex fermions, at each time step $\Delta t, 2 \Delta t, \ldots$, and for each site $n_i$ ($i = 1, 2, \ldots, N$), the probability of projective measurement being applied is $p_m \in (0,1)$. 
Instead of tracking the quantum trajectory, we focus on the two-point correlation function $C_{ij}(t) \equiv \langle c_i^{\dagger} c_j \rangle_t$. 
Under the unitary dynamics $U_t$ from $t$ to $t + \Delta t$, the correlation function evolves as $C_{ij}(t + \Delta t) = [U_t^* C(t) U_t^{\rm T}]_{ij}$. 
The projective measurement on site $n_m$ updates $C(t + \Delta t)$ to $C(t + \Delta t + 0^+)$ as follows~\cite{bravyi2004b}. 
If the measurement outcome is $n_m = 1$ with the probability $\langle n_m \rangle_{t + \Delta t}$, the correlation function is updated as
\begin{equation}
    C_{ij}(t + \Delta t + 0^+) = \delta_{im} \delta_{jm} + (1 - \delta_{im})(1 - \delta_{jm}) \left[ C_{ij}(t + \Delta t) - \frac{C_{im}(t + \Delta t) C_{mj}(t + \Delta t)}{\langle n_m \rangle_{t + \Delta t}} \right] \, ;
\end{equation}
if the measurement outcome is $n_m = 0$ with the probability $1 - \langle n_m \rangle_{t + \Delta t}$, the update is
\begin{equation}
    C_{ij}(t + \Delta t + 0^+) = (1 - \delta_{im})(1 - \delta_{jm}) \left[ C_{ij}(t + \Delta t) + \frac{C_{im}(t + \Delta t) C_{mj}(t + \Delta t)}{1 - \langle n_m \rangle_{t + \Delta t}} \right]\, .
\end{equation}
The eigenvalues $\xi_i$'s of the correlation matrix $C$ give the $\alpha$-R\'enyi entropy $S_{\alpha} = \sum_{i=1}^N g_{s \alpha}(\xi_i)$ with $ g_{s \alpha}(\xi) \equiv  \ln\left[ \xi^{\alpha} + (1 - \xi)^{\alpha} \right]/(1 -\alpha)$~\cite{cheong2004}.

For the monitored dynamics of Majorana fermions, we also track the two-point correlation function $D_{ij}(t) \equiv \ii \langle [\psi_i^{\dag} ,\psi_j]/2 \rangle_t$. At each time step and for each pair $ \ii \gamma_{2i-1}\gamma_{2i}$ ($i = 1,2,\ldots, N$), the probability of measurement being applied is $p_m \in (0,1)$.
Under the unitary dynamics $O_t$ from $t$ to $t + \Delta t$, $D_{ij}(t)$ is updated to $D_{ij}(t + \Delta t ) = [O_t D(t) O_t^{\rm T}]_{ij}$. 
The projective measurement on $ \ii \gamma_{2m-1}\gamma_{2m}$ updates $D(t + \Delta t )$ to $D(t + \Delta t + 0^+)$ as follows~\cite{bravyi2004b}. 
If the measurement outcome is $\ii \gamma_{2m-1}\gamma_{2m} =  1$ with the probability $( 1 + \langle \ii \gamma_{2m-1}\gamma_{2m} \rangle_{t + \Delta t} )/2$, the correlation function is updated as
\begin{equation}
    D_{ij}(t + \Delta t + 0^+) = \begin{cases}
        \delta_{i,2m-1} \delta_{j,2m} - \delta_{i,2m} \delta_{j,2m - 1} & (i \in \{2m-1, 2m \}, j \in \{2m-1, 2m \}), \\
        D_{ij}(t + \Delta t) + \frac{D_{i1}(t + \Delta t) D_{2j}(t + \Delta t) - D_{i2}(t + \Delta t) D_{1j}(t + \Delta t) }{1 +  \langle \ii \gamma_{2m-1}\gamma_{2m} \rangle_{t + \Delta t} }  & (i \notin \{2m-1, 2m \}, j \notin \{2m-1, 2m \}) , \\
        0 & (\text{otherwise}); 
    \end{cases}
\end{equation}
if the measurement outcome is $\ii \gamma_{2m-1}\gamma_{2m} =  -1$ with the probability $( 1 - \langle \ii \gamma_{2m-1}\gamma_{2m} \rangle_{t + \Delta t} )/2$, the update is
\begin{equation}
    D_{ij}(t + \Delta t + 0^+) = \begin{cases}
        -\delta_{i,2m-1} \delta_{j,2m} + \delta_{i,2m} \delta_{j,2m - 1} & (i \in \{2m-1, 2m \}, j \in \{2m-1, 2m \}), \\
        D_{ij}(t + \Delta t) - \frac{D_{i1}(t + \Delta t) D_{2j}(t + \Delta t) - D_{i2}(t + \Delta t) D_{1j}(t + \Delta t) }{1 -  \langle \ii \gamma_{2m-1}\gamma_{2m} \rangle_{t + \Delta t} }  & (i \notin \{2m-1, 2m \}, j \notin \{2m-1, 2m \}), \\
        0 & (\text{otherwise}).
    \end{cases}
\end{equation}
Due to $D = - D^{\rm T}$, the eigenvalues of $D$ appear in $(\lambda_i, - \lambda_i)$'s ($i = 1,2,\ldots,N$) pairs, giving $S_{\alpha} = \sum_{i=1}^N g_{s \alpha}((1 + \lambda_i)/2)$~\cite{vidal2003}.

\subsection{Additional numerical results and parameters}

\begin{figure}[htb]
  \centering
  \includegraphics[width=1\linewidth]{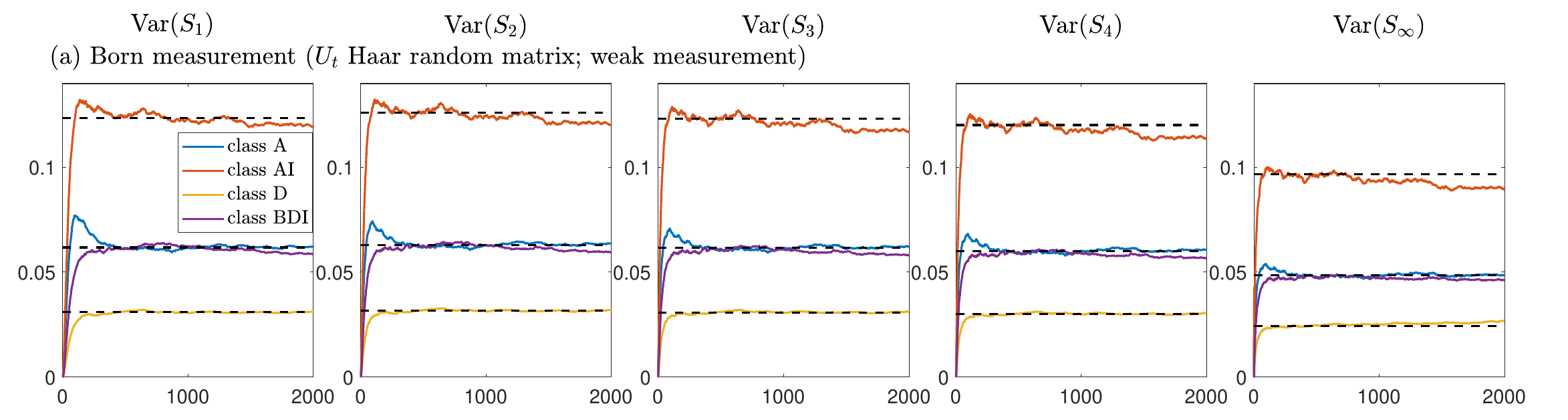}
  \includegraphics[width=1\linewidth]{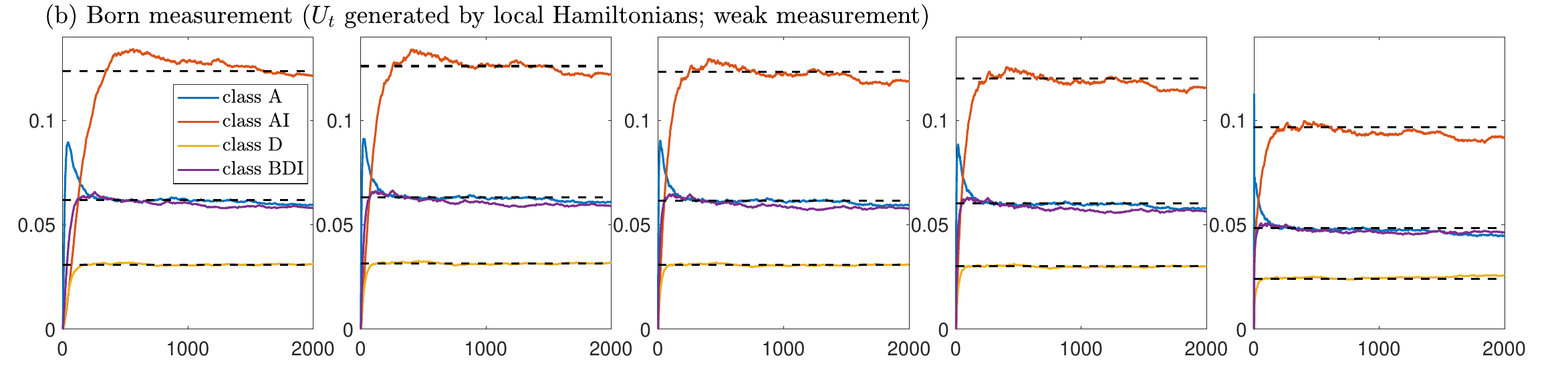}
  \includegraphics[width=01\linewidth]{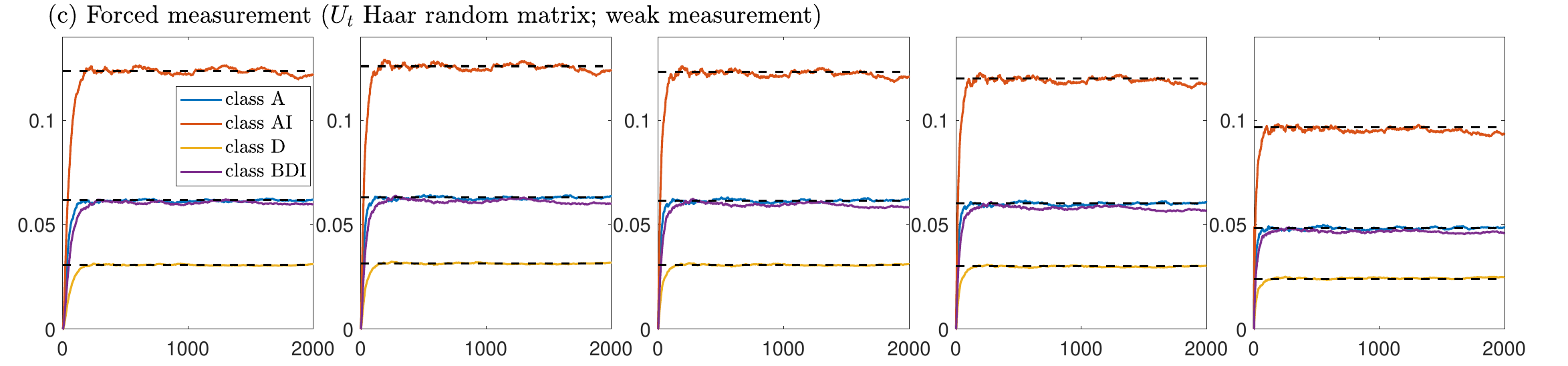}
  \includegraphics[width=1\linewidth]{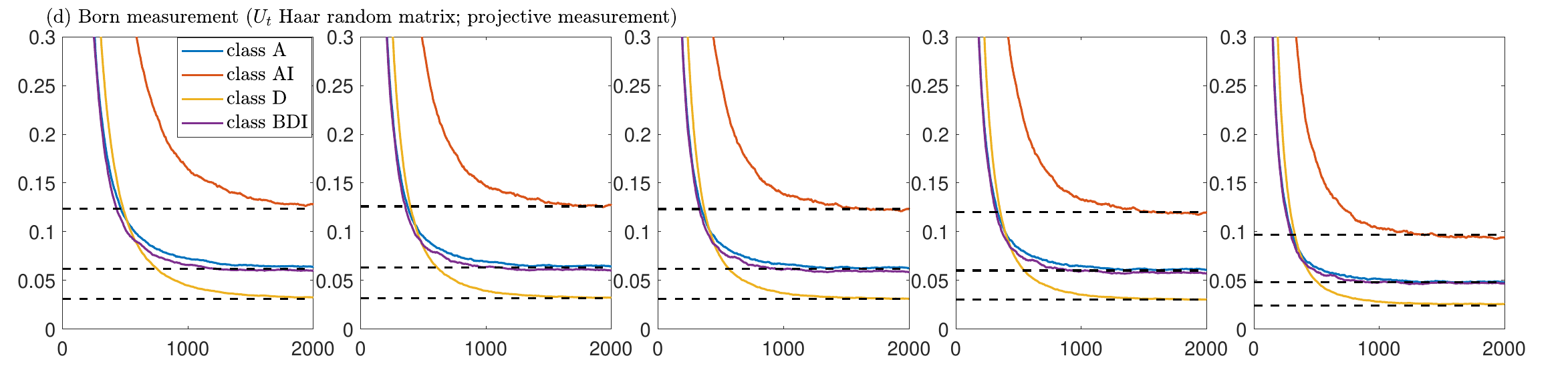} 
 \caption{Variance ${\rm Var}(S_{\alpha})$ of entropy in different types of non-unitary dynamics.
Each column corresponds to different $\alpha = 1,2,3,4$, and $\infty$ (see the top). 
The dashed lines are the analytical values [Eqs.~(\ref{eq: app varS}) and (\ref{eq: app varS2})].
(a)~Weak Born measurement and $U_t$ being a Haar-random matrix. 
(b)~Weak Born measurement and $U_t$ generated by the Hamiltonians [Eqs.~(\ref{eq: app h class A}) and (\ref{eq: app h class D})]. 
(c)~Weak forced measurement and $U_t$ being a Haar-random matrix. 
(d)~Projective measurement and $U_t$ being a Haar random matrix. 
See the parameters in the text.}
\label{fig app: var S}
\end{figure}

\begin{figure}[bt]
  \centering
  \includegraphics[width=1\linewidth]{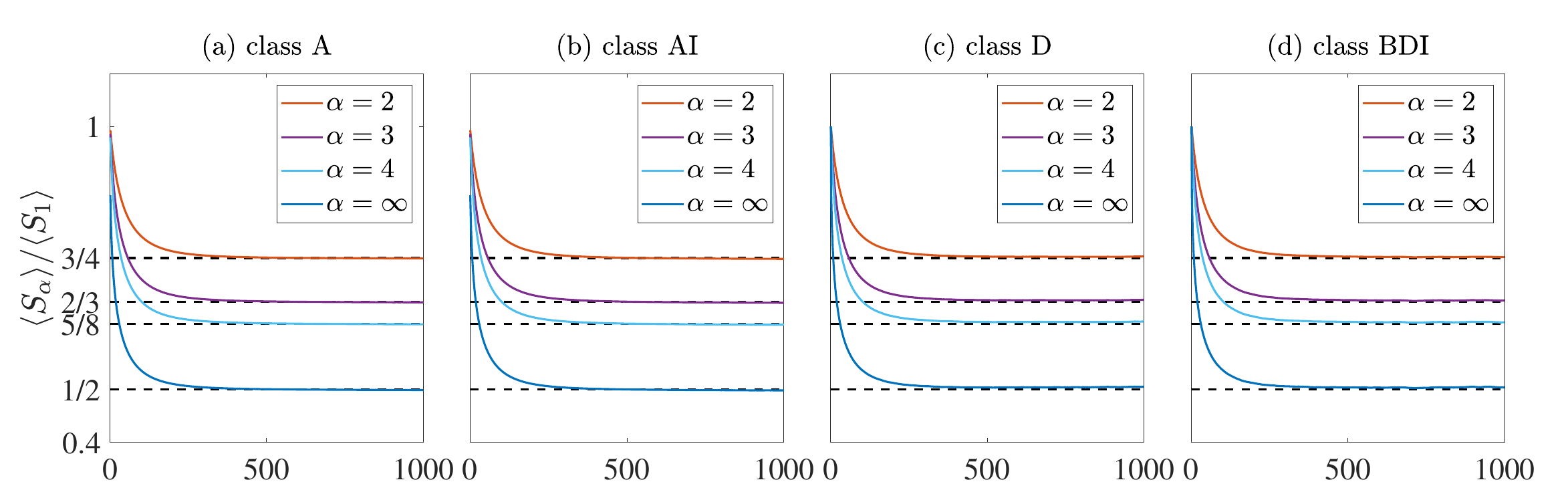}
  \caption{Simulation of Born measurement with $U_t$ being a Haar-random matrix for (a)~class A, (b)~class AI, (c)~class D, and (d)~class BDI.
  The ratios between the R\'enyi entropy $\langle S_{\alpha} \rangle$ ($\alpha = 2, 3, 4, \infty$) and von Neumann entropy $\langle S_1 \rangle$ are shown as a function of time.
  See the parameters in the text.}
  \label{fig app: S_alpha ratio}
\end{figure}

\begin{figure}[bt]
  \centering
  \includegraphics[width=0.6\linewidth]{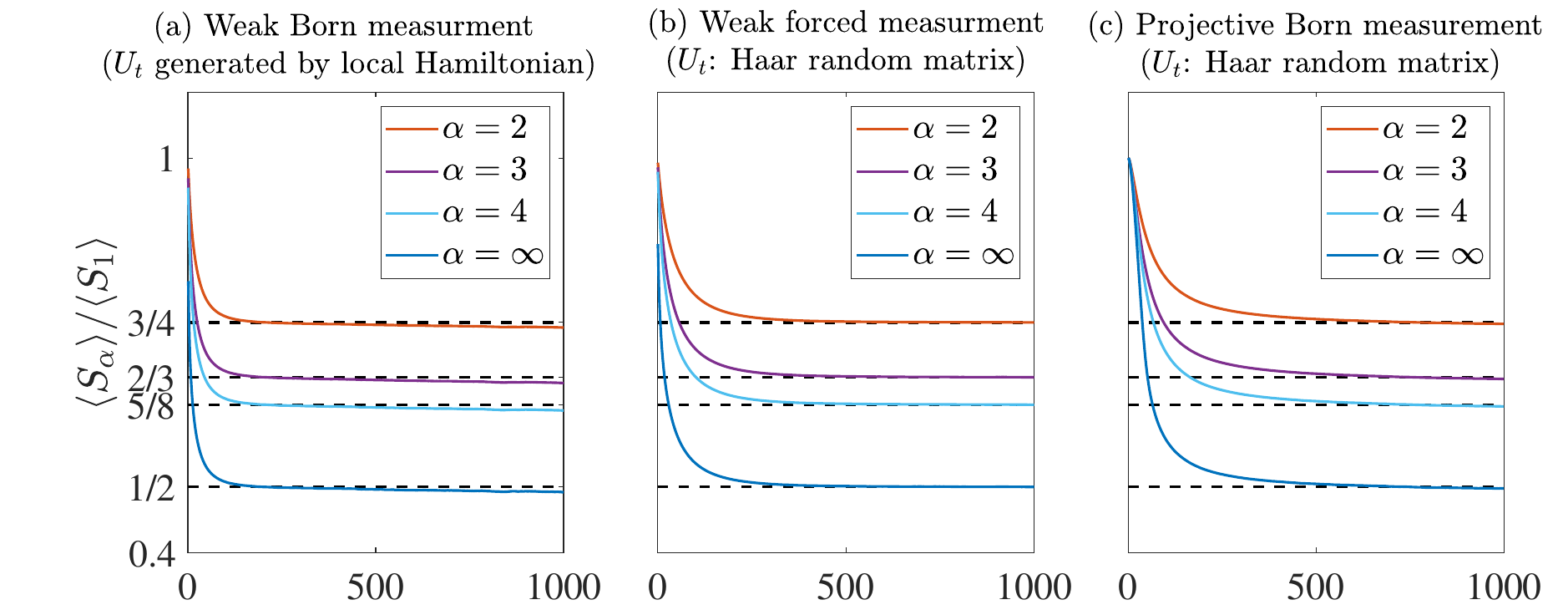}
  \caption{Simulation of different types of non-unitary dynamics with $L_{\rm eff}$ in class A.
  The ratios between the R\'enyi entropy $\langle S_{\alpha} \rangle$ ($\alpha = 2, 3, 4, \infty$) and von Neumann entropy $\langle S_1 \rangle$ are shown as a function of time.
  See the parameters in the text.}
  \label{fig app: S_alpha ratio models}
\end{figure}

\begin{figure}[hbt]
  \centering
  \includegraphics[width=1\linewidth]{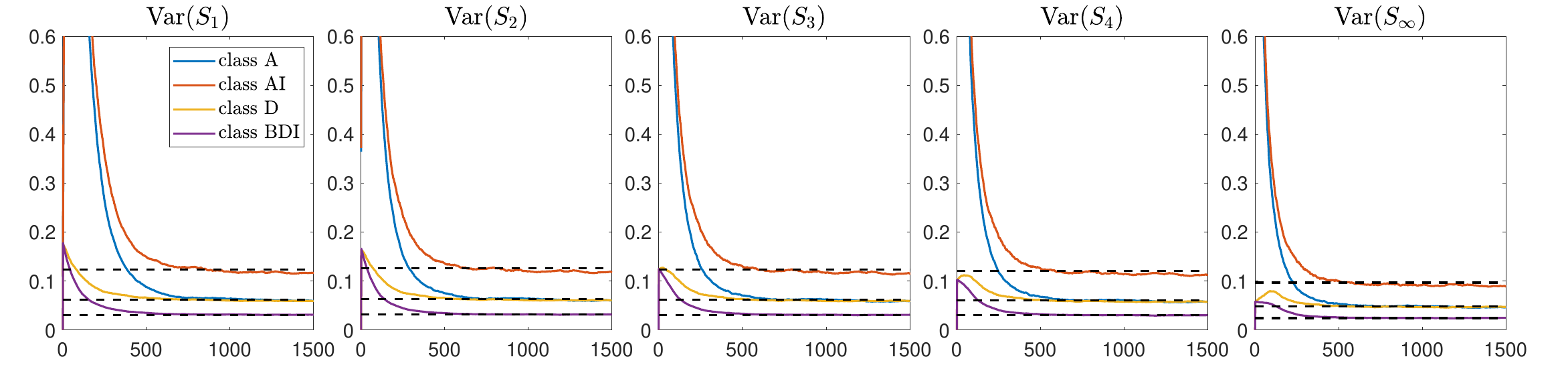}
  \caption{Variance ${\rm Var}(S_{\alpha})$ of entropy in monitored dynamics beginning with finite-temperature density matrices.
  Each column corresponds to different $\alpha = 1,2,3,4$, and $\infty$ (see the top).  
  The dashed lines are the analytical values.
  See the parameters in the text.}
  \label{fig app: S_rho0}
\end{figure}

To demonstrate the universality of our results, we simulate the monitored dynamics with the unitary dynamics $\mU_t = e^{\ii \mH_t \Delta t}$, where $\mH_t = \sum_{i,j} c_i^{\dag} (H_t)_{ij} c_j$ is a quadratic Hamiltonian with short-range hopping. 
Note that we consistently use $\mH_t$  to denote the second-quantized Hamiltonian while we use $H_t$ to denote its single-particle counterpart and analyze symmetry of $H_t$. 
We consider $\mH_t$ in a 2D $L_x \times L_y$ square lattice:
\begin{equation} \label{eq: app h class A}
  \mH_t = \sum_{\br} \sum_{\mu = x,y} J_{\br,e_{\mu} }(t)c_{\br + e_{\mu}}^{\dag} c_{\br} + {\rm H.c.} \, ,
\end{equation}
where $c_{\br}$ is the fermionic annihilation operator at site ${\bm r}$. 
(i) If $J_{\br,e_{\mu} }(t)$'s are independent complex Gaussian variables with zero mean and variance $ \langle J_{\br,e_{\mu} }(t) J^*_{\br^{\prime},e_{\nu} }(t) \rangle = 2J^2 \delta_{\br,\br^{\prime}} \delta_{\mu,\nu} \delta_{t,t^{\prime}} $, the single-particle Hamiltonian $H_t$ does not respect any symmetry other than Hermiticity and therefore belongs to symmetry class A.
(ii) If we have $\ii J_{\br,e_{\mu} }(t) \in \mathbb{R}$ and $ \langle J_{\br,e_{\mu} }(t) J^*_{\br^{\prime},e_{\nu} }(t) \rangle = J^2 \delta_{\br,\br^{\prime}} \delta_{\mu,\nu} \delta_{t,t^{\prime}} $, $H_t$ satisfies $H_t^{\rm T} = - H_t$ and hence belongs to class D.

We also consider a quadratic Majorana Hamiltonian $\mH_t = \sum_{i,j} \psi_i (H_t)_{ij} \psi_j$ with short-range hopping.
We consider a 2D $L_x \times L_y$ square lattice, and there are two Majorana operators $\psi_{\br}^{1}$ and $\psi_{\br}^2$ on each site $\br$. 
The Hamiltonian is given as 
\begin{equation} \label{eq: app h class D}
  \mH(t) = \frac{\ii}{2} \sum_{\br} \sum_{\mu = x,y} \sum_{i,j=1,2}  J^{i,j}_{\br,e_{\mu} }(t) \psi_{\br+e_{\mu}}^{i} \psi_{\br}^{j} \, .
\end{equation}
(i)~If $J^{i,j}_{\br,e_{\mu} }(t)$'s are independent real Gaussian variables with zero mean and variance $\langle J^{i,j}_{\br,e_{\mu} }(t) J^{m,n}_{\br,e_{\nu} }(t^{\prime}) \rangle = \delta_{i,m} \delta_{j,n} \delta_{\mu,\nu}, \delta_{t,t^{\prime}}J^2 $, the single-particle Hamiltonian $H_t$ satisfies $H_t^{\rm T} = -H_t$ and hence belongs to class D.
(ii)~If we have $J^{1,2}_{\br,e_{\mu} }(t) = J^{2,1}_{\br,e_{\nu} }(t) = 0$, and other $J^{i,j}_{\br,e_{\mu} }(t)$'s are independent real Gaussian variables with zero mean and variance $J^2$, besides $H_t^{\rm T} = -H_t$, $H_t$ satisfies $\sigma_z H_t \sigma_z = H_t$ with $\sigma_z$ being the Pauli matrix acting on the basis $(\psi_{\br}^{1},\psi_{\br}^{2})$. 
Consequently, $H_t$ is diagonalized into two blocks, both of which belong to class D.  

In Figs.~2 (a), (c), and (d) of the main text, we simulate the complex fermions under forced measurement using the discrete formalism (Sec.~\ref{subsec: weak m}).
The unitary dynamics $U_t$ is either a Haar-random matrix or generated by the Hamiltonian $H_t$ in Eq.~(\ref{eq: app h class A}) with the complex hopping. 
The parameters are $J = 1$, $\Delta t = 1$, $\sqrt{\gamma} = 0.4$, and $L_x \times L_y = N \times 1$ ($9 \geq N \geq 4$). 
In Fig.~2 (b) of the main text, we simulate the complex fermions under Born measurement (Sec.~\ref{subsec: weak m}). 
The unitary operator $U_t$ is a Haar-random matrix, and the measurement strength is $\sqrt{\gamma} = 0.4$.
For each parameter, we simulate at least $10^4$ realizations.

In Fig.~3 of the main text, we simulate Born measurement using the discrete formalism (Sec.~\ref{subsec: weak m}).
The unitary operator $U_t$ is distributed uniformly in the Haar measure with required symmetry.
When $L_{\rm eff}$ belongs to classes A and AI, the number of complex fermions is $N = 200$, and the measurement strength is $\sqrt{\gamma} = 0.2$.
When $L_{\rm eff}$ belongs to classes D and DIII, the number of Majorana fermions is $2N = 200$, and the measurement strength is $\sqrt{\gamma} = 0.05$.
For each parameter, we simulate at least $10^4$ realizations.

In Fig.~\ref{fig app: var S}\,(a), we simulate the same Born measurements as Fig.~3 of the main text and show ${\rm Var}(S_{\alpha})$ with different $\alpha$ ($\alpha = 1,2,3,4,\infty$). 
In Fig.~\ref{fig app: var S}\,(b), we simulate Born measurement and $U_t = e^{\ii H_t \Delta t}$ with $H_t$ in Eqs.~(\ref{eq: app h class A}) and (\ref{eq: app h class D}).
The parameters are $\Delta  t  =1$, $J = 1$, and $L_x = L_y = 8$ for all the symmetry classes. 
The measurement strength is $\sqrt{\gamma} = 0.3$ when $L_{\rm eff}$ belongs to class A, 
$\sqrt{\gamma} = 0.1$ when $L_{\rm eff}$ belongs to classes AI, 
and $\sqrt{\gamma} = 0.05$ when $L_{\rm eff}$ belong to classes D and DIII. 
In Fig.~\ref{fig app: var S}\,(c), we simulate forced measurement with $U_t$ being a Haar-random matrix.
When $L_{\rm eff}$ belongs to classes A and AI, the number of complex fermions is $N = 100$ and $400$, respectively, and the measurement strength is $\sqrt{\gamma} = 0.2$.
When $L_{\rm eff}$ belongs to classes D and BDI, the number of Majorana fermions is $2N = 200$, and the measurement strength is $\sqrt{\gamma} = 0.05$.
In Fig.~\ref{fig app: var S}\,(d), we simulate projective Born measurement (see Sec.~\ref{subsec: projective m}) with $U_t$ being a Haar-random matrix.
When $L_{\rm eff}$ belongs to classes A and AI, the number of complex fermions is $N = 100$ and $400$, respectively, 
and the measurement probability is $p_m = 0.02$ and $0.017$, respectively. 
When $L_{\rm eff}$ belongs to classes D and BDI, the number of Majorana fermions is $2N = 200$, and the measurement probability is $p_m = 0.02$ and $0.015$, respectively.
For each parameter, we simulate at least $10^4$ realizations.

The models and parameters used in Fig.~\ref{fig app: S_alpha ratio} are the same as those in Fig.~\ref{fig app: var S}\,(a). 
The models and parameters used in Fig.~\ref{fig app: S_alpha ratio models}\,(a), (b), and (c) are the same as those in Figs.~\ref{fig app: var S}\,(b), (c), and (d), respectively.

To further substantiate the universality, we simulate the monitored dynamics beginning with finite-temperature density matrices within the discrete formalism (Sec.~\ref{subsec: weak m}).
For complex fermions (Majorana fermions), we choose the initial density matrix as
\begin{equation}
\rho_0 = \prod_{i=1}^N e^{- n_i/2} \quad \left( \rho_0 = \prod_{i=1}^N e^{- \ii  \psi_{2i-1} \psi_{2i}} \right).
\end{equation}
The other setup and parameters of the dynamics are identical to those in Fig.~3 of the main text and Fig.~\ref{fig app: var S}.
As shown in Fig.~\ref{fig app: S_rho0}, 
the fluctuations of the R\'enyi entropy first increase and then reduce to the predicted universal values.

\bibliography{ref}